\documentclass[manuscript]{aastex6}

\fullcollaborationName{The Friends of AASTeX Collaboration}

\begin{document}

%% LaTeX will automatically break titles if they run longer than
%% one line. However, you may use \\ to force a line break if
%% you desire.

\title{Formation of the Unequal-Mass Binary Protostars in L1551\,NE by Rotationally-Driven Fragmentation}

%% Use \author, \affil, plus the \and command to format author and affiliation 
%% information.  If done correctly the peer review system will be able to
%% automatically put the author and affiliation information from the manuscript
%% and save the corresponding author the trouble of entering it by hand.
%%
%% The \affil should be used to document primary affiliations and the
%% \altaffil should be used for secondary affiliations, titles, or email.

%% Authors with the same affiliation can be grouped in a single
%% \author and \affil call.
\author{Jeremy Lim}
\affil{Department of Physics, The University of Hong Kong, Pokfulam Road, Hong Kong \\
\& \\
Laboratory for Space Research, Faculty of Science, The University of Hong Kong, Pokfulam Road, Hong Kong}

\author{Tomoyuki Hanawa}
\affil{Center for Frontier Science, Chiba University, Inage-ku, Chiba 263-8522, Japan}

\author{Paul K. H. Yeung}
\affil{Department of Physics, The University of Hong Kong, Pokfulam Road, Hong Kong}

\author{Shigehisa Takakuwa}
\affil{Department of Physics and Astronomy, Graduate School of Science and Engineering, Kagoshima University, 1-21-35 Korimoto, Kagoshima, Kagoshima, 890-0065, Japan}

\author{Tomoaki Matsumoto}
\affil{Faculty of Humanity and Environment, Hosei University, Chiyoda-ku, Tokyo 102-8160, Japan}

%% Use the \and command so offset the last author.
\and

\author{Kazuya Saigo}
\affil{Department of Physical Science, Graduate School of Science, Osaka Prefecture University, 1-1 Gakuen-cho, Naka-ku, Sakai, Osaka 599-8531, Japan}

%% Notice that each of these authors has alternate affiliations, which
%% are identified by the \altaffilmark after each name.  Specify alternate
%% affiliation information with \altaffiltext, with one command per each
%% affiliation.

%% Mark off the abstract in the ``abstract'' environment. 

\begin{abstract}
We present observations at 7\,mm that fully resolve the two circumstellar disks, and a reanalyses of archival observations at 3.5\,cm that resolve along their major axes the two ionized jets, of the class\,I binary protostellar system L1551\,NE.  We show that the two circumstellar disks are better fit by a shallow inner and steep outer power-law than a truncated power-law.  The two disks have very different transition radii between their inner and outer regions of $\sim$18.6\,AU and $\sim$8.9\,AU respectively.  Assuming that they are intrinsically circular and geometrically thin, we find that the two circumstellar disks are parallel with each other and orthogonal in projection to their respective ionized jets.  Furthermore, the two disks are closely aligned if not parallel with their circumbinary disk.  Over an interval of $\sim$10\,yr, source B (possessing the circumsecondary disk) has moved northwards with respect to and likely away from source A, indicating an orbital motion in the same direction as the rotational motion of their circumbinary disk.  All the aforementioned elements therefore share the same axis for their angular momentum, indicating that L1551\,NE is a product of rotationally-driven fragmentation of its parental core.  
%The much larger circumprimary disk in this system permits a rare investigation of the radial intensity profile of a circumstellar disk around a protostar.  We find that this disk, as well as the circumsecondary disk, is poorly fit by a continuous power-law, but must be truncated or exhibit a steep drop in intensity at an outer radius.  
Assuming a circular orbit, the relative disk sizes are compatible with theoretical predictions for tidal truncation by a binary system having a mass ratio of $\sim$0.2, in agreement with the reported relative separations of the two protostars from the center of their circumbinary disk.  The transition radii of both disks, however, are a factor of $\gtrsim$1.5 smaller than their predicted tidally-truncated radii.
\end{abstract}

%% Keywords should appear after the \end{abstract} command. 
%% See the online documentation for the full list of available subject
%% keywords and the rules for their use.
\keywords{(stars:) binaries (including multiple): close; (stars:) binaries: visual; (stars:) circumstellar matter; stars: jets; stars: protostars; stars: formation}

%% From the front matter, we move on to the body of the paper.
%% Sections are demarcated by \section and \subsection, respectively.
%% Observe the use of the LaTeX \label
%% command after the \subsection to give a symbolic KEY to the
%% subsection for cross-referencing in a \ref command.
%% You can use LaTeX's \ref and \label commands to keep track of
%% cross-references to sections, equations, tables, and figures.
%% That way, if you change the order of any elements, LaTeX will
%% automatically renumber them.

%% We recommend that authors also use the natbib \citep
%% and \citet commands to identify citations.  The citations are
%% tied to the reference list via symbolic KEYs. The KEY corresponds
%% to the KEY in the \bibitem in the reference list below. 

\section{Introduction} \label{sec:intro}

Fragmentation -- the internal break-up of a core into two or more fragments -- is the leading contender for how the majority of multiple stars form \citep[e.g., review by][]{Goodwin2007}.  Fission and capture, two other hypotheses considered for the formation of multiple star systems, have been ruled out or are disfavored in large part based on theoretical considerations.  Two different mechanisms have been proposed to drive fragmentation: (i) bulk (large-scale ordered) rotation; and (ii) local (small-scale) turbulence  \citep[for a succinct description of how these mechanisms lead to fragmentation, see][and also $\S\ref{sec:fragmentation}$]{Lim2016}.  Depending on the circumstances involved, these two mechanisms can predict very different geometries and dynamics for the resulting binary system: i.e., alignment between the circumstellar disks and/or spin axes of the binary components, as well as alignment between their circumstellar disks and orbital plane or between their spin and orbital axes.  Comparisons between binary properties and model predictions for their formation, however, are complicated by possible internal or external interactions during or after the protostellar phase.  Depending on the nature of the interaction, the binary system can be driven either towards or away from alignment, altering its original geometry and dynamics thus masking its formation process.

Recently, we showed that the geometrical and dynamical relationship between the binary (protostellar) system and its surrounding bulk envelope (remnant parental core) provide the crucial distinction between the two possible modes of fragmentation \citep{Lim2016}.  In the Class\,I system L1551\,IRS\,5, we found that the circumstellar disks of the binary protostars are not just closely parallel with each other, but also closely parallel with their surrounding flattened envelope.  Furthermore, the protostars are orbiting each other in the same direction as the rotation of their surrounding envelope.  The close relationship between all these different elements indicates that their angular momenta share a common axis, and points to large-scale ordered rotation for driving the fragmentation of the L1551\,IRS\,5 parental core.  Orbital solutions to measurements of the relative proper motion between the binary protostars, omitting solutions for which their circumstellar disks are predicted to be tidally truncated to sizes smaller than are observed, favour a circular or low-eccentricity orbit tilted by up to $\sim$25\degr\ from the circumstellar disks.  If the fragments that gave rise to the binary protostars in L1551\,IRS\,5 were produced at different heights or on opposite sides of the midplane in the flattened central region of a rotating core, the resulting protostars would then exhibit circumstellar disks parallel with each other and their surrounding flattened envelope but tilted from the orbital plane, as is observed.  Early during their formation, tidal interactions between the individual protostars and their surrounding, much more massive, flattened envelope would have naturally given rise to an essentially circular orbit, which has presumably been (largely) preserved during the subsequent evolution (growth) of the binary protostars.

Here, we present observations that spatially resolve for the first time the circumstellar disks of the binary protostars in the Class\,I system L1551\,NE.  Lying in the close vicinity of L1551\,IRS\,5, L1551\,NE is surrounded by a circumbinary disk \citep{Takakuwa2012,Takakuwa2014}, which itself is embedded in a flattened infalling envelope \citep{Takakuwa2013}.  The circumbinary disk exhibits clear deviations from Keplerian motion that we successfully modelled as the action of gravitational torques from the central binary system \citep{Takakuwa2014}.  These torques force material in opposing segments of the circumbinary disk to orbit faster and collide with material upstream that is orbiting more slower, resulting in a two-armed spiral pattern (comprising material compressed to higher density) imprinted onto the circumbinary disk.  At opposing segments between the two spiral arms, torques from the binary prototellar system force material to orbit slower, resulting in inflows through the circumbinary disk.  Successfully reproducing the observed spatial-kinematic structure of the circumbinary disk, the model assumes a coplanar binary system having an orbital motion in the same sense as the rotation of the circumbinary disk.  In addition, based on the projected separation of the two protostars from the inferred kinematic center of the circumbinary disk, the model asopts a binary mass ratio of 0.19.  The results presented here confirm that L1551\,NE is indeed a coplanar binary system, indicate an orbital motion for the binary protostars in the same sense as the rotational motion of their circumbinary disk, and provide entirely independent evidence in support of the inferred mass ratio of the binary system.

This manuscript is organized as follows.  Our observations and data reduction are described in $\S\ref{sec:obs}$.  To study the relative proper motion of the binary protostars, we reduced previously published archival data on the ionized jets in L1551\,NE dating back nearly 20\,yrs before our observations, as described also in $\S\ref{sec:obs}$.  The results from all these data are presented in $\S\ref{sec:results}$.  In $\S\ref{sec:CDs}$, we describe how we determined the physical parameters of the individual circumstellar disks.  In $\S\ref{sec:orbits}$, we present the relative proper motion of the binary protostars.  In $\S\ref{sec:discussion}$, we assemble all the available evidence (including that in the published literature) to infer the manner in which L1551\,NE formed.  In $\S\ref{sec:summary}$, we provide a thorough summary of our results, analyses, and interpretation. Throughout this manuscript, we assume a distance to L1551\,NE of 140\,pc \citep{Kenyon1994,Bertout1999}.

\section{Observations}\label{sec:obs}
During our observations of L1551\,IRS\,5 with the Jansky Very Large Array (VLA) on 2012 November 16, 28, and 29 as reported in \citet{Lim2016}, we also observed L1551\,NE.  The observations of L1551\,NE were interleaved with those of L1551\,IRS\,5; i.e., employing the scan sequence J0431$+$1731(the secondary calibrator) $\rightarrow$ L1551\,IRS\,5 $\rightarrow$ J0431$+$1731 $\rightarrow$ L1551\,NE $\rightarrow$ J0431$+$1731 $\rightarrow$ L1551\,IRS\,5 $\rightarrow$ J0431$+$1731 $\rightarrow$ L1551\,NE, etc.  The observations spanned a total duration of $\sim$2.5\,hr on each day.  To mitigate against rapid changes in absorption and refraction by the Earth's atmosphere, causing rapid fluctuations in the measured visibility amplitude and phase of the target source, we switched between L1551\,NE and the nearby quasar J0431$+$1731 every 20\,s.  As a check of the quality of the amplitude and phase corrections, we performed similar observations of a quasar lying close to L1551\,NE, J0431$+$2037, every $\sim$30\,mins.  This quasar also was used to check the pointing accuracy of the antennas, a task performed every $\sim$1\,hr.  The bright quasar J0510+1800 served as the bandpass calibrator, and the quasar 3C48 as the flux calibrator.

We edited, calibrated, and made maps from the data using the Common Astronomy Software Applications (CASA) package.  Considerable effort went into weeding out radio-frequency interference (RFI), which can be very weak and difficult to find, to ensure that the actual data used for making the maps is as free of contamination as is possible.  The calibration was performed in the standard manner (e.g., examples in https://casaguides.nrao.edu/index.php/Karl\_G.\_Jansky\_VLA\_Tutorials) as recommended by the observatory.  Maps were made using three different weighting schemes, natural (i.e., equal weights on all visibilities), $\rm Robust = 0.5$, and $\rm Robust = -0.25$ (robust utilize unequal weights designed to provide a more uniform sampling in $uv$-space), to accentuate different features of interest.  The synthesized beams and root-mean-square (rms) noise fluctuations ($\sigma$) of the maps thus made are summarised in Table\,\ref{tab:map parameters}.  Notice that the synthesized beams obtained using the different weighting schemes are close to circular, making it easier to visually interpret as well as to analyze the maps.   All subsequent analyses of the images obtained were made using the Astronomical Image Processing System (AIPS) and GALFIT \citep{Peng2002,Peng2010} software packages.

For studying the relative proper motion of the binary protostars, we reduced data taken by \citet{Rodriquez1995} on 1994 Apr 10 and 22, and \citet{Reipurth2002} on 2000 Nov 26--29, using also the VLA but at a wavelength of 3.5\,cm.  %The observations by \citet{Rodriquez1995} were the first to suggest that L1551\,NE might be a binary system, as was subsequently confirmed by \citet{Reipurth2002}.  Both these observations reveal free-free emission from ionized bipolar outflows (jets) associated with the binary protostars.  
We edited, calibrated, and made maps from the 1994 and 2000 observations (combining the data taken in each year) using AIPS.  The synthesized beams and root-mean-square (rms) noise fluctuations ($\sigma$) of the maps, both made with natural weighting, are summarised in Table\,\ref{tab:map parameters}.  In the observation of \citet{Rodriquez1995} in 1994, the telescope was pointed at L1551\,IRS5.  L1551\,NE is located at an angular distance of 2\farcm5 from L1551\,IRS5, almost at  the half-power point of the telescope primary beam (full-width half-maximum, FWHM, of 5\farcm3 at 3.5\,cm) in that observation.   The map made was therefore corrected for the primary beam response of the antennas.  In the observation of \citet{Reipurth2002} in 2002, the telescope was pointed at L1551\,NE.

{In all subsequent analyses, the quoted uncertainties in flux densities correspond to statistical uncertainties only, and thus do not include any systematic uncertainties (which are difficult to quantify) that arise in transferring the flux density of the primary calibrator to the secondary calibrator, and from the secondary calibrator to the target source.}

\section{Results}\label{sec:results}

\subsection{Ionized Jets}\label{sec:results jets}
\begin{figure}
\center
%\vspace{-8cm}
\epsscale{1.33}
\plotone{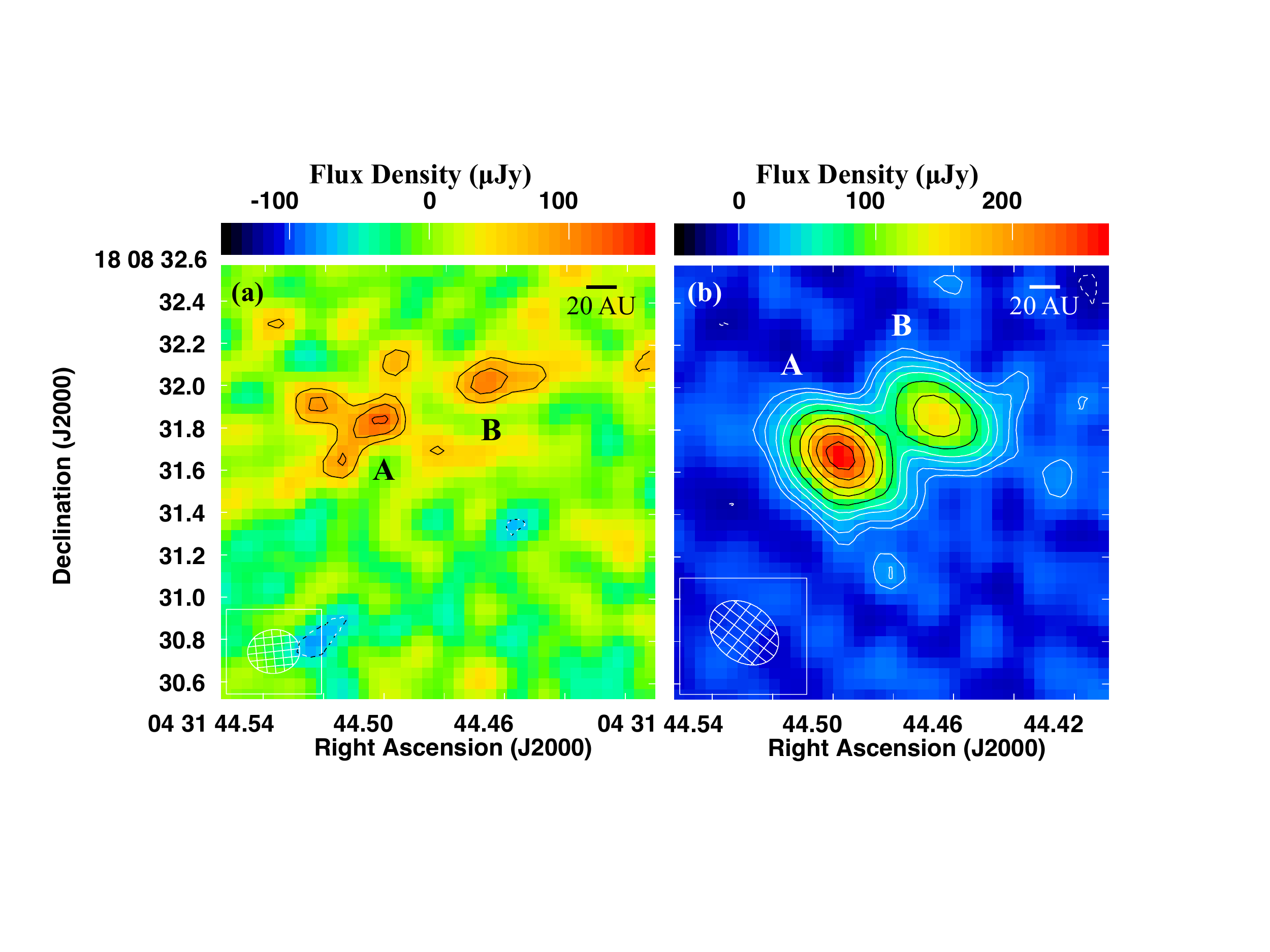}
\vspace{-3.5cm}
\renewcommand{\baselinestretch}{1.0}
\caption{3.5-cm images showing a pair of ionized jets from L1551\,NE in (a) 1994 and (b) 2002.  The brighter source to the south-east is referred to as source A, and the dimmer source to the north-west as source B \citep{Reipurth2002}.  Contour levels are plotted at $-3$, $-2$, 2, 3, and $5 \times \sigma$ (where $\sigma = 30 {\rm \ {\mu}Jy/beam}$, the rms noise level) in panel (a) and $-3$, $-2$, 2, 3, 5, 7, 10, 15, 20, and $25 \times \sigma$ ($\sigma = 8 {\rm \ {\mu}Jy/beam}$) in panel (b).  {The synthesized beam is indicated by the thatched ellipse at the lower left corner of each panel, and has a full-width half maximum (FWHM) of $\rm 264.8 \,mas \times 207.0 \,mas$ ($\rm 34.6\,AU \times 29.0\,AU$) and a position angle of $96\arcdeg.36$ in panel (a), and $\rm 356.1 \,mas \times 269.0\,mas$ ($\rm 49.9\,AU \times 37.7\,AU$) and a position angle of $52\arcdeg.1$ in panel (b).}}
\label{fig:jets}
\end{figure}

Figure\,\ref{fig:jets} shows images of L1551\,NE at 3.5\,cm made from data taken by \citet{Rodriquez1995} in 1994 (Fig.\,\ref{fig:jets}$a$) and  \citet{Reipurth2002} in 2002 (Fig.\,\ref{fig:jets}$b$).  Two sources are detected in both maps: the stronger source, located to the south-east, was referred to by \citet{Reipurth2002} as source A, and the weaker source to the north-west as source B.  We henceforth refer to these two sources in the same manner.

In Table\,\ref{tab:jets}, we list the parameters of the two sources based on a 2-dimensional Gaussian fit to each source.  In the 2002 map, which is far superior in sensitivity (a factor of nearly four lower noise) to the 1994 map, the results of the fits show that both sources are clearly resolved (at a significance level of $14\sigma$ for source A and $10\sigma$ for source B) along their major but not their minor axes.  In this map, the major axes of both sources are aligned to within measurement uncertainties (difference of $8\degr \pm 6\degr$) along an average position angle of $65\degr \pm 3\degr$ (whereas the synthesized beam has a position angle for its major axis of 52\degr.1).  We note that similar model fittings to the two sources in a robust-weighted map, which provides a higher angular resolution and hence in which the two sources are better separated, do not improve the precision of the fitting parameters.  The position angle of the ionized jets as measured at 3.5\,cm is identical to within the measurement uncertainties with the position angle of an [FeII] jet detected by \citet{Reipurth2000} and \citet{Hayashi2009} originating from the vicinity of L1551\,NE, oriented along a position angle of 63\degr\ \citep{Reipurth2000} or 64\degr\ \citep{Hayashi2009}.  Herbig-Haro objects and a bipolar molecular outflow detected in CO, all driven by L1551\,NE, lie along approximately the same position angle as the [FeII] jet \citep{Moriarty-Schieven2000}.  \citet{Reipurth2000} found that the axis of the [FeII] jet is offset from (lies to the south of) the apex of a cone-shaped nebula that is located just south-west of L1551\,NE.  This nebula comprises scattered light from the nearer side of an outflow cavity evacuated by L1551\,NE.  Based on the observed offset, \citet{Reipurth2000} attributed the [FeII] jet to source A, and associated the apex of the cone-shaped nebula with source B.

In the 1994 map, source A is formally resolved (at the 3.7$\sigma$ confidence level) but not source B (only at the 2.6$\sigma$ level).  In this map, the major axis of source A is different by $45\degr \pm 15\degr$ from that of the same source in the 2002 map.  Instead, in the 1994 map, the major axis of source A is aligned within measurement uncertainties to the major axis of the synthesized beam.  Given that L1551\,NE was located close to the half-power point of the telescope primary beam in the 1994 observation and therefore subject to both bandwidth smearing and, perhaps even more detrimentally, telescope pointing errors, we place little weight on the measured source dimensions in this map.

\subsection{Circumstellar Disks}\label{sec: results disks}
\begin{figure}
\center
%\vspace{-8cm}
\epsscale{1.33}
\plotone{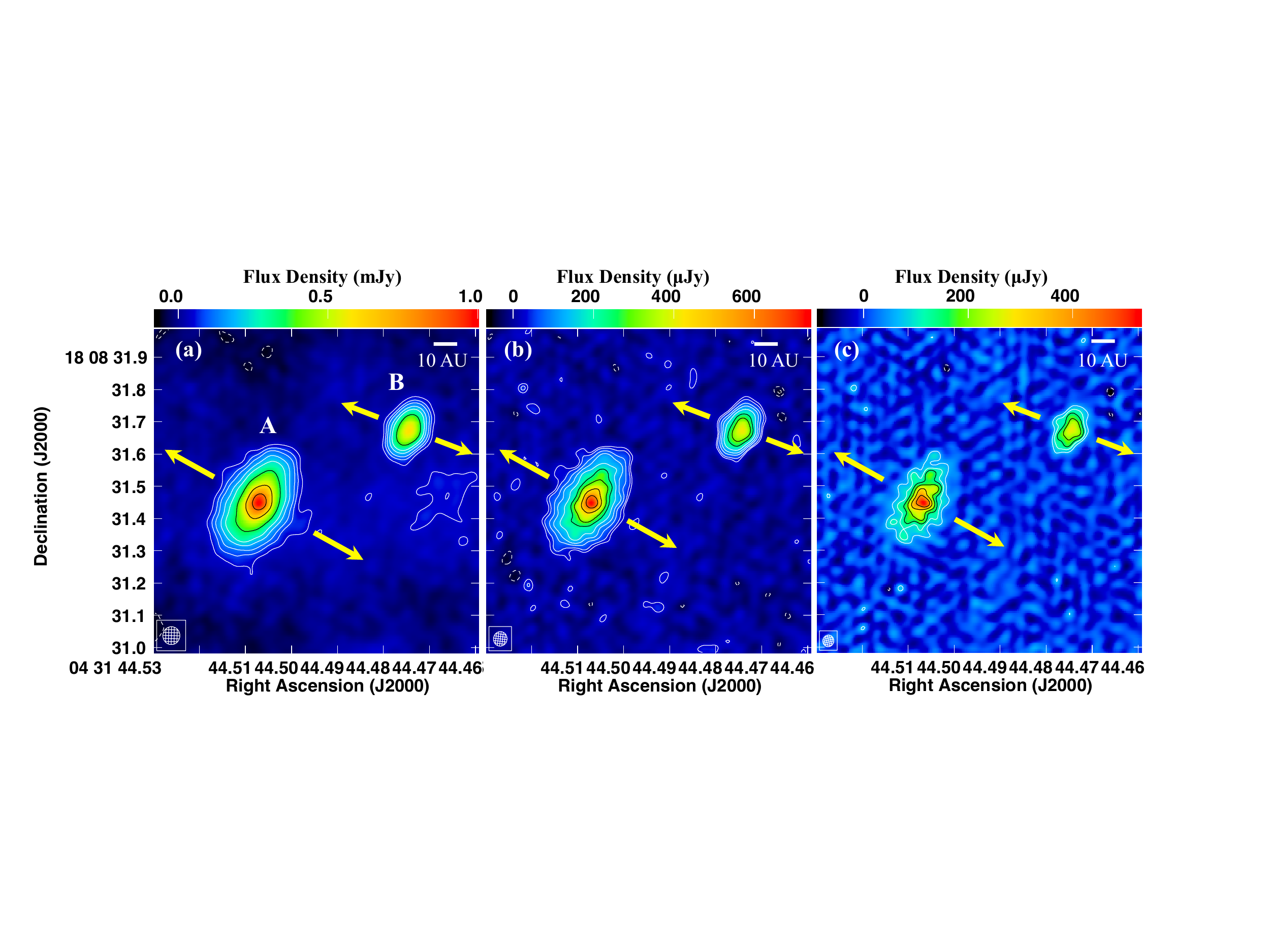}
\vspace{-4.5cm}
\renewcommand{\baselinestretch}{1.0}
\caption{7-mm images showing emission primarily from a pair of circumstellar dust disks in L1551\,NE made with (a) natural, (b) $\rm robust=0.5$, and (c) $\rm robust=-0.25$ weighting of the data.  Arrows indicate the position angles of ionized jets from Sources A and B as derived by fitting a 2-dimensional Gaussian to each source in Figure\,\ref{fig:jets}(b) (results listed in Table\,\ref{tab:jets}).  Note the weak extension of source B along the direction of its ionized jet in all three panels.  In panel (c), the central region of source A can be seen to be elongated along the direction of its ionized jet.  Contour levels are plotted at $-4$, 4, 7, 10, 15, 20, 30, 50, 70, and $90 \times \sigma$ (where $\sigma = 12 {\rm \ {\mu}Jy/beam}$, the rms noise level) in panel (a), $-3$, $-2$, 2, 3, 5, 7, 10, 20, and $25 \times \sigma$ ($\sigma = 15 {\rm \ {\mu}Jy/beam}$) in panel (b), and $-3$, 3, 5, 7, 10, 15, and $20 \times \sigma$ ($\sigma = 20 {\rm \ {\mu}Jy/beam}$) in panel (c).  The FWHM of the synthesized beam in each weighting scheme is indicated by the thatched ellipse at the lower left corner of each panel, {and has a FWHM of $\rm 55.4 \,mas \times 52.5 \,mas$ ($\rm 7.8\,AU \times 7.4\,AU$) and a position angle of $-1.20\arcdeg.36$ in panel (a), $\rm 44.9 \,mas \times 41.8 \,mas$ ($\rm 6.3\,AU \times 5.9\,AU$) and a position angle of $-6\arcdeg.5$ in panel (b), and $\rm 36.3 \,mas \times 33.8 \,mas$ ($\rm 5.1\,AU \times 4.7\,AU$) and a position angle of $-12\arcdeg.7$ in panel (c).}}
\label{fig:CDs}
\end{figure}

Figure\,\ref{fig:CDs} show our images of L1551\,NE at 7\,mm made with three different weightings, natural weighting that provides the {lowest noise level but also the poorest angular resolution of $\rm 55.4\,mas \times 52.5\,mas$ or $\rm 7.8\,AU \times 7.4\,AU$ (Fig.\,\ref{fig:CDs}$a$), {$\rm Robust=0.5$} weighting that only slightly increases the noise level but significantly improves the angular resolution to $\rm 44.9\,mas \times 41.8\,mas$ or $\rm 6.3\,AU \times 5.9\,AU$ (Fig.\,\ref{fig:CDs}$b$), and {$\rm Robust=-0.25$} weighting that provides close to the highest angular resolution possible with our data of $\rm 36.3\,mas \times 33.8\,mas$ or $\rm 5.1\,AU \times 4.7\,AU$} at the expense of a significantly higher noise level (Fig.\,\ref{fig:CDs}$c$).  A simple visual inspection reveals that both sources A and B are clearly resolved along their major and minor axes.  Source A is much larger and also has a higher peak as well as integrated flux density than source B.  Both sources are elongated in a direction perpendicular to their ionized jets (indicated by arrows in Fig.\,\ref{fig:CDs}) as traced at 3.5\,cm (Fig.\,\ref{fig:jets}$b$), and so their emission must originate primarily from dust in their circumstellar disks.  {Emission from dust at larger spatial scales, namely that in the circumbinary disk as imaged at 0.85\,mm with Submillimeter Array (ALMA) \citep{Takakuwa2012} and in follow-up observations also at 0.85\,mm with the Atacama Large Millimeter and Submillimeter Array (ALMA) \citep{Takakuwa2014}, as well as dust in the envelope around the circumbinary disk as imaged at 0.85\,mm with the SMA \citep{Takakuwa2013}, is entirely resolved out in our observation (which has a much higher angular resolution, and lacks relatively short baselines, compared with the ALMA and SMA observations).}

In all the maps shown in Figure\,\ref{fig:CDs}, source B exhibits an appreciable elongation along the north-east to south-west direction that extends beyond, and is perpendicular to the major axis of, its main body.  This elongation is aligned with its ionized jet, so that, at 7\,mm, the emission of source B along its minor axis must include a weak contribution from free-free emission associated with its ionized jet.  This situation is similar to that found for both components of L1551\,IRS5 at 7\,mm, where the emission from each source is contributed by both ionized gas and dust \citep{Lim2006,Lim2016}.  In the highest angular-resolution map at 7\,mm shown in Figure\,\ref{fig:CDs}$c$, the central peak in source A can be seen to be elongated in a direction perpendicular to its main body and aligned instead with its ionized jet.  Thus, at 7\,mm, the emission from the central region of source A must also include a contribution from free-free emission associated with its ionized jet.

\section{Physical Parameters Circumstellar Disks}\label{sec:CDs}

Because, in the images at 7\,mm, the emission from both sources include a weak contribution from their ionized jets, we first tried to remove the jets before fitting models to the disks.  Using the task IMFIT in AIPS, we started by attempting to fit a two component (one to represent the jet and the other the disk), 2-dimensional, Gaussian function to sources A and B individually in the naturally-weighted map (which provides the highest S/N ratio, and hence traces the circumstellar disks furthest out).  All such attempts either failed to converge or provided non-physical results (e.g., negative intensities for one of the components) for both sources.  This failure is in sharp contrast to our success using the same strategy for L1551\,IRS\,5, where a two component, 2-dimensional, Gaussian function provided a satisfactory fit to each of the two sources in this system at 7\,mm \citep{Lim2016}.  %From such fits to both sources in L1551\,IRS\,5, we were able to derive the inclinations (from the ratio in the dimensions of their minor to major axes) of their circumstellar disks and the position angles of their major axes.   
Below, we explain why such a model fails to fit the image of either sources in L1551\,NE.
%In the case of L\,1551\,NE, however, a 2-dimensional Gaussian function provides a poor fit to the circumstellar disk of source A, whereas the contribution from the jet in source B is too weak to provide meaningful constraints, as we will demonstrate next.

%%As we shall now show, the particularly well-resolved circumstellar disk of source A allows us to explicitly reject a Gaussian model, and more importantly to differentiate between different power-law models, for its radial intensity profile.  Because its emission includes a weak contribution from an ionized jet (seen most clearly in Fig.\,\ref{fig:CDs}$c$), we started by fitting (using the task IMFIT in AIPS) a two component (one for the ionized jet and the other for the circumstellar dust disk), 2-dimensional, Gaussian function to the naturally-weighted map (which provides the highest S/N ratio, and hence traces the circumstellar disk furthest out) of source A \citep[cf.][for the binary components in L1551\,IRS5]{Lim2016}.  All such attempts failed, with the program either failing to converge or providing non-physical results (i.e., a negative intensity for one of the components).  

\begin{figure}
\center
%\vspace{-8cm}
\epsscale{1.46}
\plotone{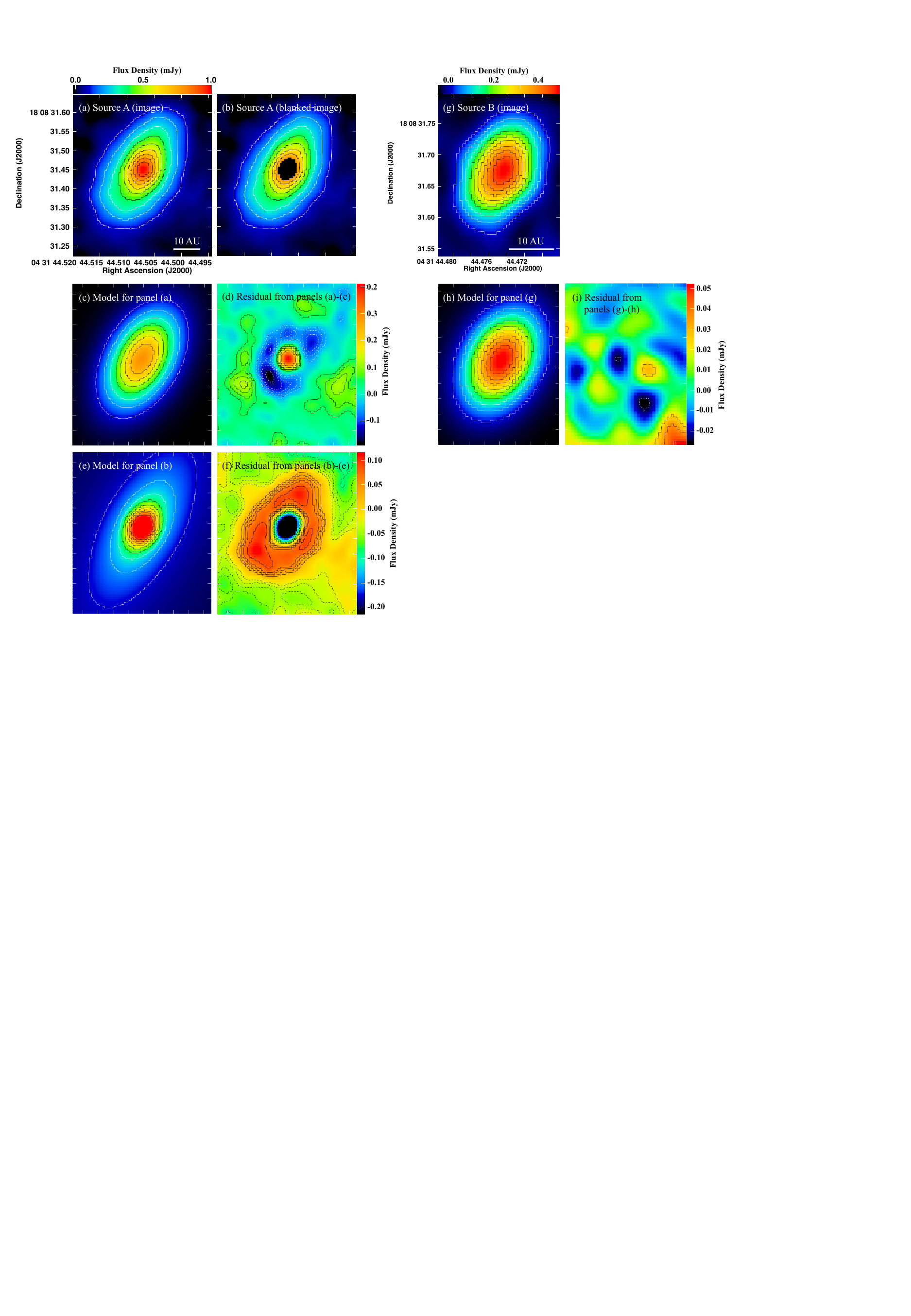}
\vspace{-17.5cm}
\renewcommand{\baselinestretch}{0.7}
\caption{2-dimensional Gaussian fits to the naturally-weighted 7-mm images of (a) source A, (b) source A with its central region having a size of approximately the FWHM of the synthesized beam blanked out, and (g) source B.  Panel (c) is the model fit to the unblanked image of source A in panel (a), and panel (d) the residuals (image$-$model).  Panel (e) is the model fit to the centrally blanked image of source A in panel (b), and panel (f) the residuals.  Panel (h) is the model fit to source B in panel (g), and panel (i) the residuals.  Colors and contours in panels (c) and (e) are the same as in panels (a)--(b), where the contour levels are plotted at 10\%, 20\%, ..., and 90\% of the peak intensity of source A, permitting a direct comparison between the model fits and the image of this source.  Similarly, colors and contours in panels (h) are the same as in panel (g), where the contour levels are plotted at 10\%, 20\%, ..., and 90\% of the peak intensity of source B, permitting a direct comparison between the model fit and the image of this source.   Contour levels in residual maps plotted at $-10$, $-7$, $-5$, $-4$, $-3$, $-2$, 2, 3, 5, 7, and $10 \times \sigma$ (where $\sigma = 12 {\rm \ {\mu}Jy/beam}$, the rms noise level).  Unlike the less well resolved image of source B, the much better resolved image of source A cannot be satisfactorily fit by a 2-dimensional Gaussian function.}
\label{fig:Gaussian fits}
\end{figure}

GALFIT, unlike IMFIT (in AIPS), does not try to fit for the (spatially unresolved) central region of a source within an area spanned by the FWHM of the synthesized beam.  This feature is convenient for our purpose so as to mitigate the contribution from the ionized jet emanating from the center of each source.  (The model fitted by GALFIT therefore makes no statement about the radial intensity profile within a central area spanned by the FWHM of the synthesized beam.)  We therefore started by fitting a 2-D Gaussian function (i.e., one component only, corresponding to the circumstellar disk) to source A in the naturally-weighted map.  Figure\,\ref{fig:Gaussian fits}$c$ shows the resulting best-fit model {(reduced-$\chi^2 = 6.50$)}.  This model can be directly compared with the image of source A shown in Figure\,\ref{fig:Gaussian fits}$a$, where the contour levels are plotted at the same levels in flux density (from 10\% to 90\%, in steps of 10\%, of the peak intensity of source A) and the colors span the same range in flux density (from the minimum to the maximum of the image shown in Fig.\,\ref{fig:Gaussian fits}$a$).  Figure\,\ref{fig:Gaussian fits}$d$ shows the residuals (image$-$model) from the fit.  The most prominent feature in the residual map is a strong central positive peak,
%, which presumably includes a contribution from if not dominated by the ionized jet in source A.  Encircling the central peak is a conspicuous negative ring and a fainter outer positive ring, indicating that a Gaussian function provides a poor fit to the circumstellar disk.  
encircled by a conspicuous negative ring and a fainter outer positive ring, indicating that a Gaussian function provides a poor fit to the circumstellar disk.  As a check, we blanked different sized areas at the center of source A, and fitted a 2-D Gaussian function to the remaining emission.  Figure\,\ref{fig:Gaussian fits}$b$ shows an example where the central region of source A having a size of approximately the FWHM of the synthesized beam has been blanked out.  The model fit of a 2-D Gaussian function {(reduced-$\chi^2 = 21.39$)} is shown in Figure\,\ref{fig:Gaussian fits}$e$, and the residuals in Figure\,\ref{fig:Gaussian fits}$f$.  The residual map shows a negative ring around the central blanked area and a surrounding positive ring; the same pattern is seen no matter the size of the central area blanked out up to about twice the FWHM of the synthesized beam, the largest that we tried.  Thus, the reason why a two-component, 2-dimensional, Gaussian function fails to provide a satisfactory fit to source A is because its circumstellar disk simply does not have a Gaussian radial intensity profile.  
%In all cases, however, the residual image displays a conspicuous positive ring, indicating that a Gaussian function provides a poor fit (too strongly declining) to the circumstellar disk of source A.  Figure\,\ref{fig:Gaussian fits}$b$ shows an example where the central region of source A having a size of approximately the FWHM of the synthesized beam has been blanked out.  The model fit of a 2-D Gaussian function is shown in Figure\,\ref{fig:Gaussian fits}$e$, and the residuals in Figure\,\ref{fig:Gaussian fits}$f$.

Source B is, visually, much smaller and spanned by fewer resolution elements than source A.  Unlike source A, source B can be satisfactorily fit (reduced-$\chi^2 = 1.15$) by a 2-dimensional Gaussian function (corresponding to its circumstellar disk) as shown in Figure\,\ref{fig:Gaussian fits}$h$.  The fitted Gaussian model can be directly compared with the image of source B shown in Figure\,\ref{fig:Gaussian fits}$g$, where the contour levels are plotted at the same levels in flux density (from 10\% to 90\%, in steps of 10\%, of the peak intensity of source B) and the colors span the same range in flux density (from the minimum to the maximum of the image shown in Fig.\,\ref{fig:Gaussian fits}$g$).  The residuals are shown in Figure\,\ref{fig:Gaussian fits}$i$, all of which are below $3\sigma$ within the detectable body of source B.  Thus, the failure to fit a two-component, 2-dimensional, Gaussian function to source B is because its jet is simply too weak to provide meaningful constraints.

The 2-dimensional Gaussian function fitted to source B has a FWHM along its major axis of 0\farcs99, roughly comparable to the values found by fitting 2-dimensional Gaussian functions to the two circumstellar disks in L1551\,IRS5 of 0\farcs122 and 0\farcs092 \citep[see Table\,2 of][]{Lim2016}.  The maps used for these fits have a similar FWHM for their synthesized beams of about 0\farcs055.  On the other hand, the circumstellar disk of source A, which as we show below is over twice as large as that of source B, cannot be fit by a 2-dimensional Gaussian function.  Our ability to satisfactorily fit a 2-dimensional Gaussian function to the circumstellar disk of source B, as well as to each of the two circumstellar disks in L1551\,IRS5, is likely because their radial intensity profiles are dominated by their synthesized beams (which are Gaussian function) and not because these disks actually have Gaussian radial intensity profiles.  

\begin{figure}
\center
\vspace{-1cm}
\epsscale{1.46}
\plotone{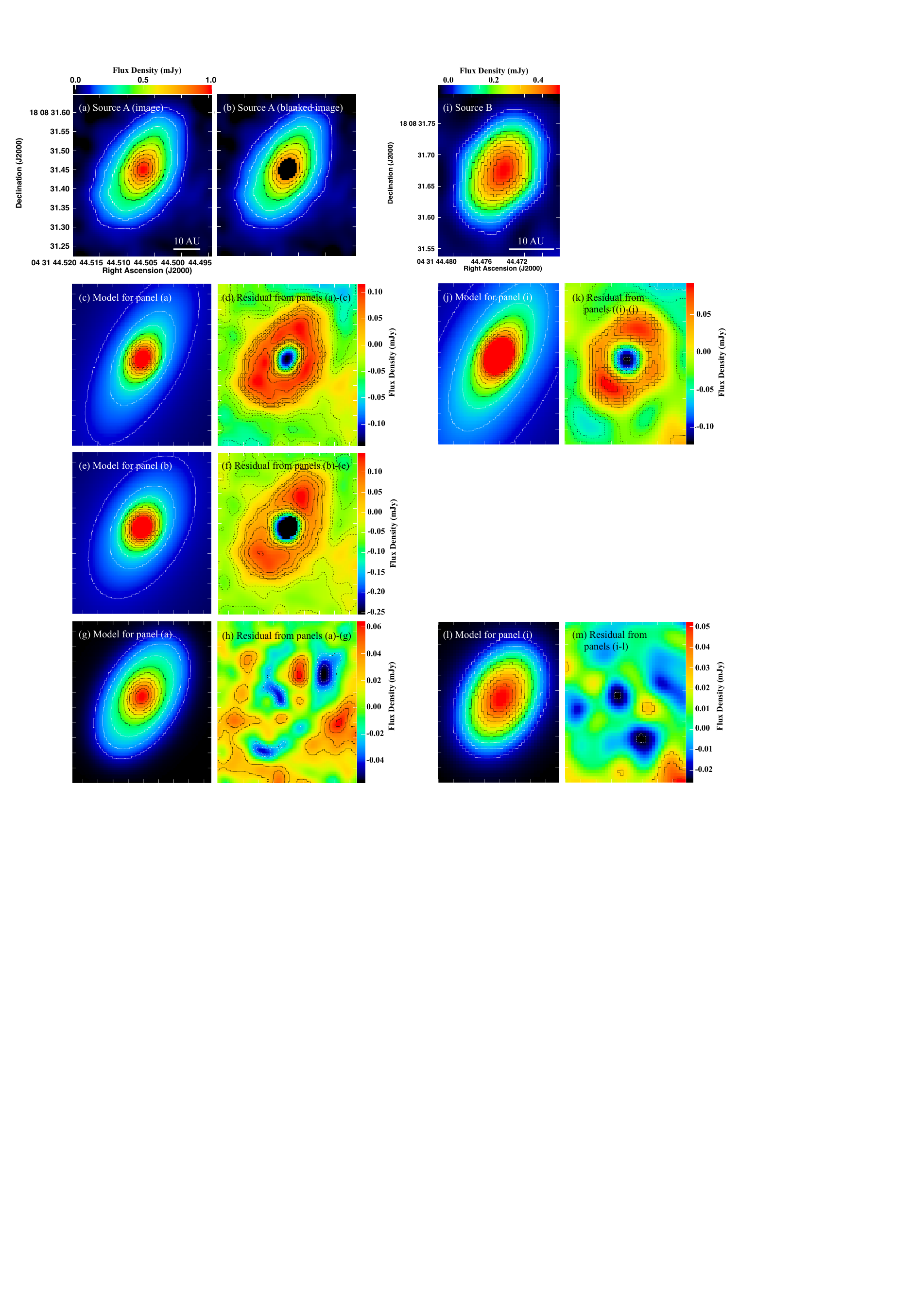}
\vspace{-13.5cm}
\renewcommand{\baselinestretch}{0.55}
\caption{Same as in Figure\,\ref{fig:Gaussian fits}, but now for 2-dimensional power-law fits.  Panel (c) is the model fit of a continuous power-law to the unblanked image of source A in panel (a), and panel (d) the residuals (image$-$model).  Panel (e) is the model fit of a continuous power-law to the centrally blanked image of source A in panel (b), and panel (f) the residuals.  Panel (g) is the model fit of a power-law that is truncated at an outer radius to the unblanked image of source A, and panel (h) the residuals.  Panel (j) is the model fit of a continuous power-law to source B in panel (i), and panel (k) the residuals.  Panel (l) is the model fit of a power-law that is truncated at an outer radius to source B, and panel (m) the residuals.  Colors and contours in panels (c), (e) and (g) are the same as in panels (a)--(b), where the contour levels are plotted at 10\%, 20\%, ..., and 90\% of the peak intensity of source A, permitting a direct comparison between the model fits and the image of this source.  Similarly, colors and contours in panels (j) and (l) are the same as in panel (i), where the contour levels are plotted at 10\%, 20\%, ..., and 90\% of the peak intensity of source B, permitting a direct comparison between the model fit and the image of this source.  Contour levels in residual maps plotted at $-10$, $-7$, $-5$, $-4$, $-3$, $-2$, 2, 3, 5, 7, and $10 \times \sigma$ (where $\sigma = 12 {\rm \ {\mu}Jy/beam}$, the rms noise level).  Both sources are poorly fit by a continous power-law, but quite well fit by a power-law that is truncated at an outer radius.}
\label{fig:PL fits}
\end{figure}

Physically-motivated models \citep[see brief review in][]{Lim2016} such as power-law profiles, designed to mimic power-law surface density and temperature profiles, having an inner as well as an outer truncation radius or taper are usually fitted to images of circumstellar disks.  Unlike optically-revealed objects for which the spectral energy distributions in the near- to mid-infrared provide constraints on a central cavity in their circumstellar disks, no such constraints are possible for protostars.  Conveniently, GALFIT does not attempt to fit for the centrally-unresolved region where a cavity might be present.  We started by fitting a 2-dimensional power-law, with no outer truncation, to sources A and B.  Figure\,\ref{fig:PL fits}$c$ shows the best fit of such a model {(reduced-$\chi^2 = 25.16$)} to the unblanked image of source A  shown in Figure\,\ref{fig:PL fits}$a$, and Figure\,\ref{fig:PL fits}$d$ the residuals.  Figure\,\ref{fig:PL fits}$e$ shows the corresponding model fit {(reduced-$\chi^2 = 21.39$)} and Figure\,\ref{fig:PL fits}$f$ the residuals for the centrally-blanked image of source A shown in Figure\,\ref{fig:PL fits}$b$.  In both cases, the residual map shows a negative central circular region or negative ring around the central blanked area and a surrounding positive ring, indicating that an untruncated 2-dimensional power-law provides a poor fit to the circumstellar disk of source A.  The same is true for source B, where Figure\,\ref{fig:PL fits}$j$ shows the fitted model {(reduced-$\chi^2 = 10.97$)} and Figure\,\ref{fig:PL fits}$k$ the residuals.  The residual map also shows a negative central circular region and a surrounding positive ring, just like the residual map of source A shown in Figure\,\ref{fig:PL fits}$d$.  

Given that circumstellar disks in binary systems are predicted to be truncated by tidal interactions with their neighboring companions, we then tried fitting a 2-dimensional power-law that is truncated at an outer radius.  %(Recall that GALFIT does not attempt to fit for the central spatilaly-unresolved region, and therefore makes no statement about the existence or otherwise of an inner truncation radius.)  
Figure\,\ref{fig:PL fits}$g$ shows the model fit and Figure\,\ref{fig:PL fits}$h$ residuals for the unblanked image of source A (Fig.\,\ref{fig:PL fits}$a$).  The fit is much improved (reduced-$\chi^2  = 3.58$, {versus a reduced-$\chi^2 = 6.50$ for a 2-dimensional Gaussian and a reduced-$\chi^2 = 25.16$ for a 2-dimensional power law with no outer truncation}) as reflected by the relatively weak residuals, although a faint negative ring is visible indicating a systematic deviation between the fitted model and the image.  Figure\,\ref{fig:PL fits}$l$ shows the corresponding model fit and Figure\,\ref{fig:PL fits}$m$ the residuals for the unblanked image of source B (Fig.\,\ref{fig:PL fits}$i$).  Once again, the fit is much improved (reduced-$\chi^2 = 1.16$) over an untruncated 2-dimensional power-law {(reduced-$\chi^2 = 10.97$)}, although in the case of source B providing no better a fit than a 2-dimensional Gaussian function {(reduced-$\chi^2 = 1.15$)}.

In L1551\,IRS5, a NUKER function, comprising a relatively shallow inner power-law and a very steep outer power-law (i.e., a tapered rather than a truncated profile), was fitted to the two circumstellar disks \citep{Lim2016}.  This function provides a smooth transition between the inner inner and outer power-laws, a feature that was deemed to be more physical than a discontinuous transition.  The NUKER function is parameterised as: 
\begin{equation}
I(r) = I_b \ 2^{{\beta-\gamma}\over\alpha} ({r \over r_b})^{-\gamma} [1 + ({r \over r_b})^\alpha]^{\gamma-\beta \over \alpha} \ \ \ , 
\end{equation}
where $I(r)$ is the intensity, $I$, as a function of radius, $r$, $\gamma$ is the inner power-law slope, $\beta$ the outer power-law slope, $\alpha$ controls the sharpness of the transition between the two power laws (larger $\alpha$ indicating a sharper transition), $r_b$ the break radius at which the slope is the average of $\beta$ and $\gamma$ or, equivalently, the radius of maximum curvature in logarithmic units, and $I_b$ the intensity at $r_b$.  Just like for the two circumstellar disks in L1551\,IRS5, we found that the central position, inclination (as determined from the ratio in dimensions of the minor to major axes), and position angle of the major axis of sources A and B to be essentially constant independent of $\alpha$.  As $\alpha$ increases (i.e., the transition between the inner and outer power-law becomes sharper), the break radius $r_b$ decreases somewhat and very rapidly converges.  Likewise, both $\gamma$ (the inner power-law index) and $\beta$ (the outer power-law index) also rapidly converge with increasing $\alpha$, such that  $\beta \gg \gamma$ irrespective of $\alpha$.  

\begin{figure}
\center
%\vspace{-8cm}
\epsscale{1.46}
\plotone{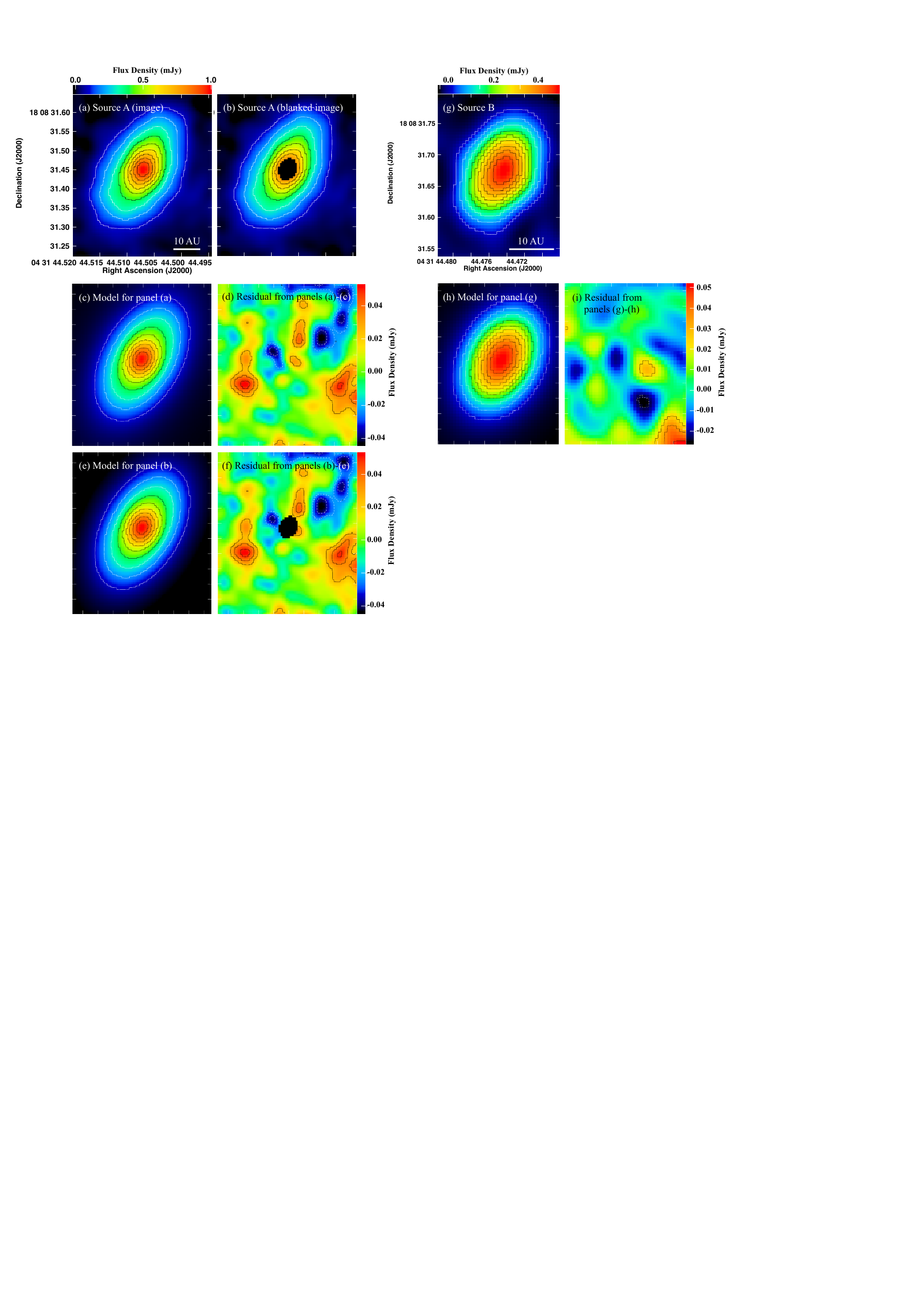}
\vspace{-17.5cm}
\renewcommand{\baselinestretch}{0.8}
\caption{Same as in Figure\,\ref{fig:Gaussian fits} and Figure\,\ref{fig:PL fits}, but now for 2-dimensional NUKER (double power-law with a smooth transition) fits.  Panel (c) is the model fit to the unblanked image of source A in panel (a), and panel (d) the residuals (image$-$model).  Panel (e) is the model fit to the centrally blanked image of source A in panel (b), and panel (f) the residuals.  Panel (h) is the model fit to source B in panel (g), and panel (i) the residuals.  Colors and contours in panels (c) and (e) are the same as in panels (a)--(b), where the contour levels are plotted at 10\%, 20\%, ..., and 90\% of the peak intensity of source A, permitting a direct comparison between the model fits and the image of this source.  Similarly, colors and contours in panels (h) are the same as in panel (g), where the contour levels are plotted at 10\%, 20\%, ..., and 90\% of the peak intensity of source B, permitting a direct comparison between the model fit and the image of this source.  Contour levels in residual maps plotted at $-10$, $-7$, $-5$, $-4$, $-3$, $-2$, 2, 3, 5, 7, and $10 \times \sigma$ (where $\sigma = 12 {\rm \ {\mu}Jy/beam}$, the rms noise level).  NUKER functions provide obviously better or statistically superior fits to both sources than either 2-dimensional Gaussian (Fig.\,\ref{fig:Gaussian fits}) or single power-law (Fig.\,\ref{fig:PL fits}) fits.}
\label{fig:NK fits}
\end{figure}

Fixing therefore the central location, inclination, and position angle of each source, we list in Table\,\ref{tab:NUKER} the other parameters of the best-fit NUKER function at the largest value of $\alpha$ for which a solution is obtainable.  In this way, we obtained a break radius of $r_b$$\sim$133\,mas ($\sim$18.6\,AU) for source A.   Figure\,\ref{fig:NK fits}$c$ shows the model fit to the unblanked image of source A (Fig.\,\ref{fig:NK fits}$a$) and Figure\,\ref{fig:NK fits}$d$ the residuals.  This fit (reduced-$\chi^2 = 2.25$) is, by far, the best among all those considered ({versus a reduced-$\chi^2 = 6.50$ for a 2-dimensional Gaussian, a reduced-$\chi^2 = 25.16$ for a 2-dimensional power law with no outer truncation, and a reduced-$\chi^2 = 3.58$ for a 2-dimensional power law truncated at an outer radius)}.  Importantly, there are no clearly apparent systematic residuals indicating a systematic deviation between the fitted model and the image; nonetheless, %although no significant residuals are present in the central region where an ionized jet is detectable (and visible in maps at higher angular resolutions), 
there are low-level residuals in the outer regions that limit the goodness of the fit.  Fitting a NUKER function to the image of source A where its central region is blanked out (Fig.\,\ref{fig:NK fits}$b$), we obtained an essentially identical model fit {(reduced-$\chi^2 = 2.24$)} as shown in Figure\,\ref{fig:NK fits}$e$ and residual map as shown in Figure\,\ref{fig:NK fits}$f$.  The position angle of the major axis thus derived for the circumstellar disk of source A is $\sim$$150\degr.9$, accurately orthogonal in projection to the position angle inferred for the axis of its ionized jet of $61\degr^{+4}_{-3}$.  The corresponding model fit for source B (reduced-$\chi^2 = 1.07$) is shown in Figure\,\ref{fig:NK fits}$h$ and the residual map in Figure\,\ref{fig:NK fits}$i$.  Like for source A, this model provides a superior fit (lower reduced-$\chi^2$) to source B than a 2-dimensional Gaussian {(reduced-$\chi^2 = 1.15$)}, {a power-law with no outer truncation (reduced-$\chi^2 = 10.97$)}, or a power-law truncated at an outer radius {(reduced-$\chi^2 = 1.16$)}.  The position angle of the major axis thus derived for the circumstellar disk of source B is $\sim$$152\degr.1$, closely orthogonal in projection to the position angle inferred for the axis of its ionized jet of $69\degr^{+4}_{-5}$.  Assuming both circumstellar disks to be circular and geometrically thin, the inclination derived for the circumstellar disk of source A is $\sim$57\degr.7 and that of source B is $\sim$58\degr.0.  Their similar inclinations and position angles for their major axes imply that the two circumstellar disks are (closely) parallel.

GALFIT does not provide uncertainties for the model fitting parameters.  As a measure of the uncertainties in the inclination and position angle for the circumstellar disk of source B, we also used IMFIT (which provides uncertainties in the model parameters) to fit a 2-dimensional Gaussian function to this source.  In this manner, we derived an inclination of $56\degr.3 \pm 3\degr.8$\,deg (GALFIT reports 58\degr.2 for a 2-dimensional Gaussian fit and 58\degr.0 for a NUKER fit) and a position angle of $154\degr.5 \pm 4\degr.6$ (GALFIT reports 152\degr.2 for a 2-dimensional Gaussian fit and 152\degr.1 for a NUKER fit) for source B.  With a position angle of $69\degr^{+4}_{-5}$ for its ionized jet (Table\,\ref{tab:jets}), the circumstellar disk of source B is, within the uncertainties, orthogonal ($85\degr.5 \pm 6\degr.8$) in projection to its ionized jet.  As mentioned earlier, source A cannot be fit by a 2-dimensional Gaussian function, and so we cannot provide corresponding uncertainties for the inclination and position angle for its circumstellar disk using this method.  Nevertheless, based on the exercise conducted for source B, the uncertainties in the inclination and position angle of the major axis derived from fitting a NUKER function to source A are probably no larger than a few degrees.

\section{Orbital Motion}\label{sec:orbits}
The 1994 observation of \citet{Rodriquez1995} was the first to show that L1551\,NE comprises two sources, as was subsequently confirmed in the 2002 observation of \citet{Reipurth2002}.  To date, only these and our observation in 2012 provide useful measurements of the relative proper motion of the binary protostars.  Table\,\ref{tab:rPM} lists the positions (repeated, for convenience, from Table\,\ref{tab:jets} for the 1994 and 2002 observations), relative separations, and relative orientations of the binary protostars on the aforementioned dates.  The position for source B is derived from a 2-dimensional Gaussian fit (using IMFIT) to the natural-weighted map of this source (as mentioned earlier, IMFIT reports uncertainties in the fitting parameters, unlike GALFIT) in Figure\,\ref{fig:CDs}$a$.  The position of source A is derived from a 2-dimensional Gaussian fit (also using IMFIT) to the ionized jet in the central region of this source in the $\rm Robust=-0.25$ image of Figure\,\ref{fig:CDs}$c$; the fitted Gaussian model shares a similar position angle ($67\degr^{+5\degr}_{-4\degr}$) as that derived for the ionized jet at 3.5\,cm ($61\degr^{+4\degr}_{-3\degr}$), but is resolved also along the minor axes suggesting a contribution from the circumstellar disk to the fit.  Note that different secondary calibrators were used in the three observations of L1551\,NE, and so the positions listed in Table\,\ref{tab:rPM} are referenced with respect to a different position in the sky in each observation.  The information listed in Table\,\ref{tab:rPM} should therefore be used with caution (i.e., the uncertainty in the position of the secondary calibrator needs to be included) for deriving the absolute proper motion of L1551\,NE  (motion of the entire system across the sky).  

\begin{figure}
\center
%\vspace{-8cm}
\epsscale{1.4}
\plotone{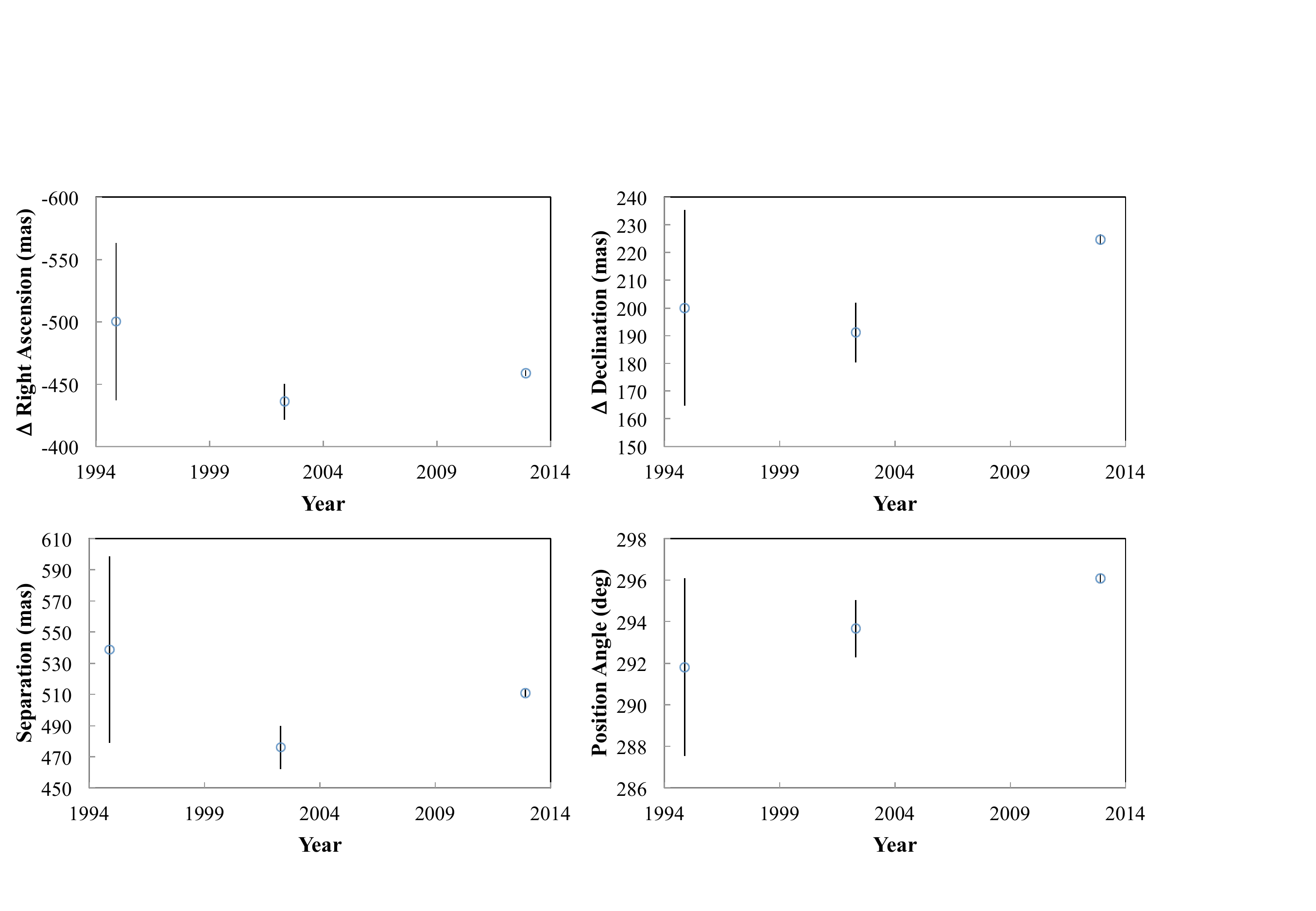}
\vspace{-0.5cm}
\renewcommand{\baselinestretch}{1.3}
\caption{Angular separation and orientation of Source B with respect to source A during the three observations reported in the text.}
\label{fig:pm}
\end{figure}

Figure\,\ref{fig:pm} shows the angular separation and orientation of source B with respect to source A over an interval spanning $\sim$18.6\,yrs.  As can be seen, there is no significant motion ({i.e., difference in positions of $\ge 3\sigma$}) of these two sources along the east-west direction, with their positions differing by $\rm 22.9\,mas \pm 14.6\,mas$ in right ascension between 2002 and 2012.  On the other hand, between 2002 and 2012, source B has moved northwards with respect to source A by $\rm 33.6\,mas \pm 11.0\,mas$ (a significance level of $3.1\sigma$).  Furthermore, source B is likely moving away (at a significance level of $2.5\sigma$) from source A.  The uncertainties in the measured source positions in 1994 are too large to detect any corresponding motion in source B between this and the later observations.

\section{Discussion}\label{sec:discussion}
The circumstellar disks of the binary protostars in L1551\,NE are parallel to each other within measurement uncertainties of a few degrees.  The close alignment between the circumstellar disks of binary protostars, however, does not by itself discriminate between different models for the formation of these systems.  Even in those systems where the binary protostars are born with misaligned circumstellar disks, tidal interactions between the protostars can align their circumstellar disks with the orbital plane \citep{Lubow2000,Bate2000}.  In such situations, tidal interactions induce disk precession; viscosity in the disks acts on the shearing motion to dissipate energy, gradually aligning the disks with the orbital plane.  \citet{Bate2000} find that such dissipative processes can align protostellar disks and their orbital plane on timescales of order 20 orbital periods, which for binary systems with a total mass of $\sim$1\,M$_\sun$ and an orbital separation of $\sim$100\,AU correspond to an interval of just $\sim$$10^4$\,yrs.  Instead, as pointed out by \citet{Lim2016} and demonstrated for the binary protostellar system L1551\,IRS5, the geometrical and dynamical relationship between the binary system and its surrounding bulk envelope provide the crucial distinction between different fragmentation models.

%The ionized jets from the two protostars in L1551\,NE are aligned to within their measurement uncertainties of $8\degr \pm 6\degr$.  At 3.5\,cm, the ionized jet of source A is stronger than that of source B, as is the case at 7\,mm where the jet of source A contributes much more appreciably to the emission in the $\rm Robust=-0.25$ map than the jet of source B.

%The two circumstellar disks in L1551\,NE are accurately (to within the measurement uncertainties) orthogonal, in projection, to their respective ionized jets.  Furthermore, their essentially identical inclinations and position angles demonstrate that the two circumstellar disks are accurately parallel with each other.

\subsection{Relationship with Circumbinary Disk}\label{sec:relationship circumbinary disk}
\citet{Takakuwa2012} inferred the bulk properties of the circumbinary disk in L1551\,NE by fitting a circular and geometrically-thin disk exhibiting Keplerian motion to channel maps in C$^{18}$O(3-2) as measured with the SMA.  This simple model reproduces the global velocity behavior of the circumbinary disk, and provides best-fit parameters of $62\degr^{+25\degr}_{-17\degr}$ for its inclination and $167\degr^{+23\degr}_{-27\degr}$ for the position angle of its major axis.

In observations at a higher angular resolution and sensitivity with the ALMA, \citet{Takakuwa2014} found clear deviations from Keplerian rotation in the circumbinary disk as measured also in C$^{18}$O(3-2).  They were able to reproduce these deviations by including gravitational torques from the binary protostars, assumed to have a circular coplanar orbit, but otherwise retained the geometry inferred by \citet{Takakuwa2012} for the circumbinary disk.  Based on the angular separation of the two sources from the inferred dynamic center of the circumbinary disk, they found a binary mass ratio of $m_B/m_A = 0.19$, where $m_A$ is the mass of the protostar corresponding to source A and $m_B$ that corresponding to source B.  From the measured orientation of sources A and B and the assumed circular coplanar orbit, \citet{Takakuwa2014} inferred an orbital separation of $\sim$145\,AU for the binary system.

The inclination and position angle for the major axis of the circumbinary disk in the model proposed by \citet{Takakuwa2012} agree, to within their measurement uncertainties, with the corresponding values we derived for the circumstellar disks (Table\,\ref{tab:NUKER}).  Although the uncertainties in these parameters for the circumbinary disk are much larger than the uncertainties in the corresponding parameters for the circumstellar disks, we note that their formal values agree to within $\sim$5\degr\ in inclination and $\sim$15\degr\ in position angle.  Thus, the circumstellar disks are not only parallel with each other, but also closely (if not accurately) parallel with their surrounding circumbinary disk.

Assuming that the equatorial plane of the circumbinary disk is orthogonal to the outflow cavity so that its eastern side is the near side, \citet{Takakuwa2012, Takakuwa2014} find that the circumbinary disk is rotating in an anticlockwise direction.  For a coplanar binary system with a circular anticlockwise orbit, at their present orbital locations source B should be moving primarily northwards and somewhat eastwards with respect to source A, increasing in angular separation \cite[see Fig.\,10 of][]{Takakuwa2014}.  The northward motion and likely increasing separation that we measure for source B with respect to source A, but smaller (no detectable) motion along the east-west direction ($\S\,\ref{sec:orbits}$), are therefore consistent with an orbital motion for the binary protostars in the same manner as the rotational motion of their surrounding circumbinary disk.

\subsection{Binary Mass Ratio}\label{sec:mass ratio}
If the binary protostars in L1551\,NE have a mass ratio of $\sim$0.19 and are in a circular orbit separated by $\sim$145\,AU as in the model described by \citet{Takakuwa2014}, the circumstellar disk of source A is predicted to be tidally truncated at a radius of $\sim$$58.4$\,AU and that of source B at a radius of $\sim$$23.5$\,AU \citep[derived from the calculations provided in][]{Pichardo2005}.  By comparison, the inferred break radius ($r_b$) is $\sim$18.6\,AU for the circumstellar disk of source A and $\sim$8.9\,AU for the circumstellar disk of source B (Table\,\ref{tab:NUKER}), both a factor of $\sim$3 smaller than their predicted tidally-truncated radii.  Equating their break radii with their tidally-truncated radii, then for a circular orbit, the predicted binary mass ratio is $\sim$0.23 and the orbital separation $\sim$47\,AU.  The binary mass ratio (which, for a given orbital eccentricity, solely determines the tidally-truncated sizes of their constituent circumstellar disks) thus inferred is closely comparable to that inferred by \citet{Takakuwa2014} of $\sim$0.19.  In this case, however, the predicted orbital separation is much smaller than the observed angular separation between the binary components in L1551\,NE of $71.5 \pm 0.4$\,AU.  Thus, the binary components of this system cannot simultaneously have a circular orbit and break radii for their circumstellar disks corresponding to their tidally-truncated radii.  %Because the ratio in the break radii of the circumstellar disks is so similar to the ratio predicted for their tidally-truncated radii for the inferred binary mass ratio of $\sim$0.2 and a circular orbit, it seems to us more likely that the break radii of both circumstellar disks are related in the same specific manner to (i.e., the same fraction of) their tidally-truncated radii rather than the binary system having a highly eccentric orbit.

In L1551\,IRS5, the relative proper motion of the binary protostars have been measured with sufficient precision to make an exploration of orbital solutions meaningful \citep{Lim2016}.  For circular orbits with orbital separations of up to $\sim$100\,AU, the (roughly comparable) break radii of the two circumstellar disks in this system can be closely comparable (somewhat smaller than) or at worse within a factor of $\sim$2 of their predicted tidally-truncated radii.  Thus, either the circumstellar disks of the binary protostars in both L1551\,IRS5 and L1551\,NE do not extend to their tidally-truncated radii, or observations at 7\,mm do not trace the overall extents of these disks.  {Observations at shorter wavelengths, where the dust emissivity is larger and hence the dust emission stronger, may better define the overall extents of these circumstellar dust disks.  Furthermore, such observations can reveal any dependence in disk sizes with wavelength, as has been found for the Class\,0 source Per-emb-14 (also known as NGC\,1333\,IRAS\,4C) and the pre-main-sequence (Classical T\,Tauri) star AS209.  The circumstellar disk of Per-emb-14 is much smaller (by a factor of about three) at 8\,mm \citep{Segura-Cox2016} than at 1.3\,mm \citep{Tobin2015}.  Similarly, for AS209, the measured size of its circumstellar disk decreases towards longer wavelengths, a behavior attributed to the radial drift of dust grains \citep{Perez2012}.}  Alternatively, the orbit of both L1551\,IRS5 and L1551\,NE may be highly eccentric, although \citet{Lim2006} found that even a moderate orbital eccentricity is highly unlikely in the case of L1551\,IRS5.  %Here, we note that the ratio in the predicted tidally-truncated radii of the two circumstellar disks in L1551\,NE is $\sim$0.40.  By comparison, the ratio of their break radii is $r_b(\rm B) / r_b(\rm A) \sim 0.48$, where $r_b(\rm B)$ is the break radius of the circumstellar disk for source B and $r_b(\rm A)$ the break radius of the circumstellar disk for source A, a ratio close to that predicted for their tidally-truncated radii for a mass ratio of $\sim$0.19.  More precisely, if indeed their measured break radii are related in a specific manner to (i.e., a certain fraction of) their tidally-truncated radii, and the orbit is circular, then for a ratio in their tidally-truncated radii of $\sim$0.48 the binary mass ratio is $\sim$0.23, close to that inferred by \citet{Takakuwa2014} of $\sim$0.19.  

\subsection{Collimated Outflows}
As mentioned in $\S\ref{sec:results jets}$, \citet{Reipurth2000} argue that the apex of a cone-shaped reflection nebula --- comprising an outflow cavity --- associated with L1551\,NE is coincident with source B, and that the [FeII] jet detected from L1551\,NE originates from source A.  To be detectable in [FeII], the jet from source A must have a higher density than that of source B at the same distance from their respective protostars (we rule out a much lower excitation for the jet from source B given that it, like the jet from source A, contains an ionized component detected in free-free emission at 3.5\,cm).  Thus, source A must have either a more powerful or a more highly collimated jet, or both (possibly related, perhaps indirectly, to the higher mass of this protostar and its larger circumstellar disk), than source B.  A more highly collimated jet provides a natural explanation for why the refection nebula has its axis passing through source B rather than source A; i.e., the walls of the outflow cavity are carved out by the poorly-collimated jet from source B.  A more highly-collimated jet also produces stronger free-free emission than a more poorly-collimated jet for the same mass-loss rate \citep{Reynolds1986}, adding to the reasons why the ionized jet from source A is brighter than that from source B.

Other scenarios cannot be ruled out, but are not supported by the available evidence or contrived.  For example, perhaps source B grew more quickly in mass (i.e., it experienced a higher accretion-rate) and drove a stronger outflow than source A, thus dominating the creation of the observed outflow cavity.  In that case, source B ought to be the more massive than source A, in contradiction with the evidence presented above.  Alternatively, the jet from source A only turned on recently and we are fortunate to be observing L1551\,NE soon after this event, an unlikely situation.

\subsection{Rotationally-Driven Fragmentation of L1551\,NE Parental Core}\label{sec:fragmentation}
Current models invoke either local (small-scale) turbulence in or the bulk (globally-ordered) rotation of cores to drive fragmentation.  In cores that have little or no bulk rotation, turbulence introduces velocity and density inhomogeneities that can seed and drive the growth of multiple density perturbations to become self gravitating \citep[e.g.,][]{Bate2002, Bate2003, Bate2005, DelgadoDonate2004a, DelgadoDonate2004b, Goodwin2004a, Goodwin2004b, Goodwin2006, Matsumoto2015}.  Multiple fragments produced in different turbulent cells are predicted to exhibit random orientations between the circumstellar disks of the binary components, and no particular relationship between the circumstellar disks and surrounding circumbinary material.  If multiple fragments are produced in a common region where turbulence conspires to create local angular momentum, however, the binary system thus assembled can exhibit quite well aligned circumstellar disks.  Nevertheless, once again, the circumstellar disks should not bear any particular relationship with their surrounding circumbinary material.

Alternatively, the large-scale ordered rotation of the core can drive dynamical instabilities to induce fragmentation during collapse.  In such models, conservation of angular momentum forces cores to become increasingly flattened as they collapse.  As a result, a disequilibrium disk-like (i.e., flattened and rotating) structure forms at the center of the core.  The central region of the core can become especially flattened if magnetic fields are invoked to direct infalling matter onto the mid-plane of the disk-like structure; the resulting structures closely resemble, at least morphologically, rotationally-supported disks, and are therefore referred to as pseudodisks \citep{Galli1993a, Galli1993b}.  By introducing an initial density or velocity perturbation, the large-scale ordered rotation of the core can drive dynamical instabilities in the form of a spiral, bar, or ring in its central flattened region \citep{Matsumoto2003, Cha2003, Machida2008}.  %The pattern of the dynamical instability does not depend on the nature of the initial velocity or density perturbations.  Furthermore, because the perturbations introduced can either promote or hinder fragmentation, in the latter case through the effective removal of angular momentum by gravitational torques from the resulting dynamical instability, increasing the amplitude of the initial perturbation does not necessarily increase the likelihood of fragmentation.  
Fragments form in localised regions of the resulting dynamical instabilities that are gravitationally unstable (according to the Toomre criterion) and have masses exceeding the local Jeans mass.  Binary prototellar systems that form through rotational fragmentation of disk-like structures should naturally exhibit a close alignment between the circumstellar disks of the binary components and the surrounding circumbinary material, and share the same sense in orbital motion.  Such a close alignment and similar sense in orbital motion is what we find for L1551\,NE, as found also by \citet{Lim2016} for L1551\,IRS5, arguing for the formation of L1551\,NE through rotational fragmentation just like in the case of L1551\,IRS5. 

In L1551\,IRS5, the circumstellar disks of the binary components have comparable sizes \citep[break radii of 12.2\,AU and 10.4\,AU respectively;][]{Lim2016}, suggesting that the binary prostars have comparable masses.  In L1551\,NE, one protostar is about five times more massive than the other.  Evidently, rotationally-driven fragmentation can lead to binary protostellar systems having either very similar or very different component masses.

\subsection{Fragmentation of L1551 Cloud}
Intriguingly, the spin axes of both the L1551\,NE and L1551\,IRS5 systems are closely oriented in space.  Specifically, the circumstellar disks of the binary protostars in L1551\,IRS5 have inclinations of $\sim$46\degr\ and position angles for their major axes of $\sim$148\degr\ \citep{Lim2016}, compared with the circumstellar disks of the binary protostars in L1551\,NE that have inclinations of $\sim$58\degr\ and position angles for their major axes of $\sim$151\degr ($\S\ref{sec:CDs}$).  Despite the close spatial orientation of their spin axes, however, the two systems exhibit opposite {senses in} spins.  Specifically, as observed from the Earth, the two components of L1551\,NE are orbiting in an anticlockwise direction ($\S\ref{sec:relationship circumbinary disk}$), whereas the two components in L1551\,IRS5 are orbiting in a clockwise direction \citep{Lim2006,Lim2016}.

In theoretical simulations, an initial density or velocity perturbation is imprinted onto rotating cores to facilitate fragmentation driven by dynamical instablities.  In molecular clouds, a ubiquitous source of perturbation is turbulence.  Different turbulent cells may have been responsible for producing the parental cores of L1551\,NE and L1551\,IRS5, and imparted on them opposite spins.  If so, then the close alignment between the spin axes of these two binary protostellar systems is purely coincidental.

L1551\,NE and L1551\,IRS5 make up one group of active star formation in the L1551 cloud.  The other is the HL\,Tau group, which comprises HL\,Tau, XZ\,Tau, LkH$\alpha$\,358, and HH\,30$^*$ (driving source of the Herbig-Haro object HH\,30).  HL\,Tau is classified as either a Class\,I object (protostar) or a Class\,II object (classical T\,Tauri star), and the others in the HL\,Tau group as Class\,II objects.  The circumstellar disk of HL\,Tau has been very well resolved with ALMA.  Based on a 2-dimensional Gaussian fit to its image, \citet{ALMA Partnership2015} derive an inclination of $46\degr.2 \pm 0\degr.2$ and a position angle for its major axis of $138\degr.2 \pm 0\degr.2$.  The ionized jet from HL\,Tau has its major axis at a position angle of $\sim$51\degr\, \citep{Mundt1990, Lopez1995, Moriarty-Schieven2006}, closely orthogonal to the circumstellar disk of this object as projected onto the sky.  Thus, surprisingly, the circumstellar disk of HL\,Tau also is closely aligned with the circumstellar disks of the binary protostars in L1551\,NE and L1551\,IRS5.  

The circumstellar disk of LkH$\alpha$\,358 has been resolved with ALMA.  Based on a 2-dimensional Gaussian fit to its image, \citet{ALMA Partnership2015} derive an inclination of $56\degr \pm 2\degr$ and a position angle for its major axis of $170\degr \pm 3\degr$.  LkH$\alpha$\,358 does not exhibit any known jet.  HH\,30$^*$ is a suspected binary based on wiggles in its optical jet \citep{Anglada2007}.  The circumbinary disk of this system has an inclination of $81\degr \pm 2\degr$ and a position angle for its major axis of $125\degr \pm 1\degr$ \citep{Guilloteau2008}.   The ionized jet from HH\,30$^*$ has its major axis at a position angle of $\sim$31\degr\, \citep{Mundt1990, Lopez1995, Moriarty-Schieven2006}, closely orthogonal to the circumbinary disk of this object as projected onto the sky.  Although distributed over a wider range of angles, the circumstellar disk of LkH$\alpha$\,358 and the circumbinary disk of HH\,30$^*$ do not appear to be randomly oriented with respect to, but instead are aligned to within a few tens of degrees of, the circumstellar disks of L1551\,NE, L1551\,IRS5, and HL\,Tau.  As a consequence, the outflows driven by all these objects, including the outflow from XZ\,Tau \citep{Krist1999, Mundt1990, Moriarty-Schieven2006} (a binary system whose circumstellar disks have not been spatially resolved), are all oriented in the north-east to south-west direction as projected onto the sky.  

If not for the counter-rotating spins of L1551\,NE and L1551\,IRS5, it would have been natural to attribute the relative close alignment between the spin axes of all the young stellar objects in the L1551 cloud to a large-scale ordered rotation of this cloud.  Instead, we note that the spin axes of all these objects are approximately orthogonal, in projection, to the major axis of the filament that comprises the L1551 cloud \citep[][whose measurements provide no evidence for any ordered rotation of this cloud]{Lin2016}; the L1551 cloud filament is itself aligned with the overall elongation of filamentary structures that make up the Taurus molecular cloud complex \citep{Mizuno1995,Goldsmith2008}.  The close alignment in the spin axes of all the young stellar objects in the L1551 cloud may therefore reflect (faster) infall and the subsequent formation of cores that are flattened along the major axis of the cloud filament.  Local turbulence may have imparted angular momentum to individual cores, thus giving rise to opposite spins between some cores.

\section{Summary and Conclusions}\label{sec:summary}

Using the VLA, we have fully resolved (i.e., along both their major and minor axes) the two circumstellar disks in the class\,I binary protostellar system L1551\,NE.  We also reanalysed archival observations at 3.5\,cm that resolve along their major axes the two ionized jets in this system.  These observations span nearly two decades, allowing us to study the relative proper motion of the binary protostars.  We found that:

\begin{itemize}

\item the stronger ionized jet of source A has a position angle for its major axis of $61\degr^{+4\degr}_{-3\degr}$, and the weaker ionized jet of source B a position angle for its major axis of $69\degr^{+4\degr}_{-5\degr}$.  Both jets are therefore aligned, as projected onto the sky, to within the measurement uncertainties (difference in position angles of $8\degr \pm 6\degr$).

\item the circumstellar disk of source A is much larger than that of source B.  The images of both circumstellar disks are better fit by a double power-law that exhibits a smooth transition between the inner and outer power-laws, than a single power-law that is abruptly truncated.  A single, untruncated, power-law is explicitly rejected for the circumstellar disks of both sources, as is a Gaussian for the circumstellar disk of source A.

\item although we find no unique solution for a double power-law fit to either circumstellar disks, the ratio of their major to minor axes as well as the position angle of their major axes do not depend on other parameters.  Assuming implicitly that the circumstellar disks are intrinsically circular and geometrically thin, we find that the circumstellar disk of source A has an inclination of 57\degr.7 and a position angle for its major axis of 150\degr.9, and the circumstellar disk of source B an inclination of 58\degr.0 and a position angle for its major axis of 152\degr.1.  With estimated uncertainties in these parameters of a few degrees, the two circumstellar disks are closely aligned if not parallel.  Furthermore, the two circumstellar disks are accurately orthogonal in projection to their respective ionized jets.

\item for the sharpest transition between the inner and outer power-laws as might be expected of tidally-truncated disks, the radius of maximum curvature in this transition is $\sim$18.6\,AU for the circumstellar disk of source A and $\sim$8.9\,AU for the circumstellar disk of source B.  Equating these transition radii with their theoretically predicted tidally-truncated radii, then, for a circular orbit, the ratio in disk sizes imply a binary mass ratio of $\sim$0.23.  This binary mass ratio is closely comparable with that inferred by \citet{Takakuwa2014} of $\sim$0.19 based on the projected angular separation between each protostar and the inferred kinematic center of the circumbinary disk.  Given the projected angular separation between the two protostars, however, the transition radii of both circumstellar disks are at least a factor of $\sim$1.5 times smaller than their predicted tidally-truncated radii if the system has a binary mass ratio of $\sim$0.2 and a circular orbit.

\item over an interval of 10\,yr, source B has moved northwards (at a significance level of $3.1\sigma$) with respect to source A.  By contrast, there is no detectable motion of these two sources along the east-west direction (significance level of only $1.6\sigma$).  Furthermore, source B is likely moving away (at a significance level of $2.5\sigma$) from source A.  All these measurements agree with the model proposed by \citet{Takakuwa2014} for the relative orientation of the two protostars at their inferred orbital locations for an anticlockwise and circular orbital motion.

\end{itemize}

The two circumstellar disks are closely aligned if not parallel not just with each other but also with their surrounding circumbinary disk, which has an inclination of $62\degr^{+25\degr}_{-17\degr}$ and a position angle for its major axis of $167\degr^{+23\degr}_{-27\degr}$.  Furthermore, the two protostars appear to be orbiting each other in same direction as the rotation of their circumbinary disk.  Both the circumstellar and circumbinary disks, as well as the orbit, of this binary system therefore share the same axes for their angular momenta, indicating that L1551\,NE formed through the rotationally-driven fragmentation of its parental core, as is the case for L1551\,IRS5 \citep{Lim2016}.  By contrast with L1551\,NE, where the two circumstellar disks have different sizes and their binary protostars different masses, the two circumstellar disks in L1551\,IRS5 have roughly comparable sizes suggesting that their binary protostars have comparable masses (for a given orbital eccentricity, the truncation radii of circumstellar disks in binary systems depend only on the binary mass ratio).  Rotationally-driven fragmentation can therefore lead to binary systems having comparable or very different component masses.

Finally, we pointed out that the circumstellar disks of the binary protostars in both L1551\,NE and L1551\,IRS5, along with their circumbinary disks or flattened circumbinary envelopes, are closely oriented in space (i.e., similar inclinations, as well as position angles for their major axes).  Indeed, all the young stellar objects in the L1551 cloud, including HL\,Tau, LkH$\alpha$358, HH30$^*$, and probably also XZ\,Tau, have spin axes that are approximately orthogonal in projection to the major axis of the filament that makes up the L1551 cloud, which itself is aligned with the major axes of the filamentary structures that make up the Taurus molecular cloud complex.  This alignment may reflect (faster) infall along and the subsequent formation of cores that are flattened across the minor axes of these filaments.  Local turbulence may have imparted angular momentum to individual cores, thus giving rise to opposite spins between some cores.

%circumstellar disks and/or outflows that are quite closely aligned in space. Yet, L1551\,IRS5 and L1551\,NE exhibit opposite spins (clockwise for L1551\,IRS5 and anticlockwise for L1551\,NE), implying that their closely-aligned spin axes cannot be attributed to large-scale ordered rotation (if any) of the L1551 cloud.  Instead, the orientation of their spin axes approximately orthogonal to the major axes of filamentary structures that make up the Taurus molecular cloud complex may reflect infall along the minor axis of these filaments followed by the formation of cores through turbulence driven fragmentation.

%% If you wish to include an acknowledgments section in your paper,
%% separate it off from the body of the text using the \acknowledgments
%% command.
\acknowledgments
We thank the anonymous referee for constructive suggestions to improve the readability of the manuscript.  J. Lim acknowledges support from the Research Grants Council of Hong Kong through grants HKU\,703512P and 17305115.  Both T. Hanawa and T. Matsumoto acknowledge the Japan Society for the Promotion of Science (JSPS) Grants-in-Aid for Scientific Research (KAKENHI) through, respectively, Grant Number JP15K05017 and 26400233.

%% To help institutions obtain information on the effectiveness of their 
%% telescopes the AAS Journals has created a group of keywords for telescope 
%% facilities. 

%% Following the acknowledgments section, use the following syntax and the
%% \facility{} macro to list the keywords of facilities used in the research 
%% for the paper.  Each keyword is check against the master list during
%% copy editing.  Individual instruments can be provided in parentheses,
%% after the keyword, but they are not verified.

\vspace{5mm}
\facilities{VLA}

\software{CASA, AIPS, GALFIT}

\newpage
\clearpage

\newpage
\clearpage

\newpage
\clearpage

\newpage
\clearpage

\newpage
\clearpage

\newpage
\clearpage

\newpage
\begin{deluxetable}{ccccccc}
\vspace{0cm}
\tabletypesize{\normalsize}
\tablecolumns{7}
\tablewidth{0pc}
\tablecaption{Map Parameters \label{tab:map parameters}}
\tablehead{
\colhead{Year} & {Wavelength} & {Map} & \colhead{\hspace{2.0cm} Synthesized Beam} & & & \colhead{rms} \vspace{-5mm} \\ 
\colhead{} & {} & {Weighting} & \colhead{\hspace{-3.5cm} Major Axis} & \colhead{\hspace{-4.5cm} Minor Axis} & \colhead{\hspace{-1.2cm} Position Angle} & \colhead{Noise} \vspace{-5mm} \\
\colhead{} & {} & {} & \colhead{\hspace{-3.5cm} (mas) (AU)} & \colhead{\hspace{-4.5cm} (mas) (AU)} & \colhead{\hspace{-1.2cm} (deg)} & \colhead{($\mu$Jy)} }
\startdata
1994.89 & 3.5\,cm & Natural & \hspace{-3.5cm} 246.8 \, 34.6 & \hspace{-4.5cm} 207.0 \, 29.0 & \hspace{-2.0cm} $96.36$ & 30 \\
2002.30 & 3.5\,cm & Natural & \hspace{-3.5cm} 356.1 \, 49.9 & \hspace{-4.5cm} 269.0 \, 37.7 & \hspace{-2.0cm} $52.10$ & 8 \\
2012.91 & 7\,mm & Natural & \hspace{-3.5cm} 55.4 \hspace{0.3cm} 7.8 & \hspace{-4.5cm} 52.5 \hspace{0.3cm} 7.4 & \hspace{-1.9cm} -$1.20$ & 12 
\vspace{-5mm} \\
        &   & Robust ($0.5$) & \hspace{-3.5cm} 44.9 \hspace{0.3cm} 6.3 & \hspace{-4.5cm} 41.8 \hspace{0.3cm} 5.9  & \hspace{-1.7cm} -$6.5$ & 15 \vspace{-5mm} \\
        &   & Robust (-0.25) & \hspace{-3.5cm} 36.3 \hspace{0.3cm} 5.1 & \hspace{-4.5cm} 33.8 \hspace{0.3cm} 4.7 & \hspace{-1.9cm} -$12.7$ & 20  \\
\enddata
%\vspace{0.1cm}
%\tablecomments{}
\end{deluxetable}

\newpage
\begin{deluxetable}{ccccccc}
\vspace{0cm}
\tabletypesize{\footnotesize}
\tablecolumns{7}
\tablewidth{0pc}
\tablecaption{Parameters Ionized Jets\label{tab:jets}}
\tablehead{
\colhead{Year} & \colhead{Source} & \colhead{Right Ascension} & \colhead{Declination} &  \colhead{Flux Density} & \colhead{Major Axis} & \colhead{Position Angle} \vspace{-0.5cm} \\
\colhead{} & \colhead{} & \colhead{(J2000)} & \colhead{(J2000)} & \colhead{($\mu$Jy)} & \colhead{(mas)} & \colhead{(deg)} \vspace{-0.5cm} \\}
\startdata
1994.89 & A & 04$\rm ^h$31$\rm ^m$44$\rm ^s$.4975$\pm 0.0038$ & $+18^\circ$08$\farcm$31\farcs90$\pm0.03$ & $269 \pm 85$ & $484^{+131}_{-138}$ & $106^{+12}_{-14}$ \vspace{-0.5cm} \\
         & B & 04$\rm ^h$31$\rm ^m$44$\rm ^s$.4625$\pm 0.0023$ & $+18^\circ$08$\farcm$32\farcs10$\pm0.02$  & $153 \pm 54$ & $252^{+97}_{-119}$ & $108^{+40}_{-41}$ \\ 
2002.30 & A & 04$\rm ^h$31$\rm ^m$44$\rm ^s$.49701$\pm 0.00048$ & $+18^\circ$08$\farcm$31\farcs673$\pm0.005$ & $348 \pm 22$ & $305^{+22}_{-20}$ & $61^{+4}_{-3}$ \vspace{-0.5cm} \\
         & B & 04$\rm ^h$31$\rm ^m$44$\rm ^s$.46643$\pm 0.00088$ & $+18^\circ$08$\farcm$31\farcs864$\pm0.009$ & $217 \pm 23$ & $359^{+34}_{-37}$ & $69^{+4}_{-5}$ \\ 
\enddata
%\vspace{0.1cm}
\tablecomments{Flux Densities listed are integrated quantities based on a two-dimensional Gaussian fit to each source.}
\end{deluxetable}
\clearpage

\begin{deluxetable}{cccccccc}
\tabletypesize{\normalsize}
\tablecolumns{8}
\tablewidth{0pc}
\tablecaption{Parameters Circumstellar Disks \label{tab:NUKER}}
\tablehead{
\colhead{Source} & \colhead{Inclination} & \colhead{Position Angle} & \colhead{\hspace{0.2cm}$\alpha$} & \colhead{\hspace{0.2cm}$\gamma$} & \colhead{\hspace{0.2cm}$\beta$} & \colhead{\hspace{1.0cm} $r_b$} \\
\colhead{} & \colhead{} & \colhead{Major Axis} & \colhead{} & \colhead{} & \colhead{} & \colhead{}  \\
%\colhead{Map} & \colhead{\hspace{3.0cm} Minimum Radius} & \colhead{\hspace{-2.0cm} } & \colhead{\hspace{-0.0cm} Inclination} & \colhead{Position} & \colhead{Brightness} \\
%\colhead{Weighting} & & & \colhead{} & \colhead{Angle} & \colhead{Temperature} \\
\colhead{} & \colhead{(deg)} & \colhead{(deg)} & \colhead{} & \colhead{} & \colhead{} & \colhead{\hspace{-0.2cm}(mas)} & \colhead{\hspace{-0.5cm} (AU)} }
\startdata
A & $57.7$ & $150.9$ & $40.0$ & $0.79$ & $4.3$ & $133.0$ & $18.6$ \\
B & $58.0$ & $152.1$ & $32.5$ & $0.45$ & $5.4$ & $63.5$ & $8.9$ \\
\enddata
%\vspace{0.1cm}
\tablecomments{NUKER fits to the {naturally-weighted images of Sources A and B (Fig.\,\ref{fig:CDs}a)} based on the largest value of $\alpha$ for which a solution is obtainable.}
\end{deluxetable}
\clearpage

\newpage
\begin{deluxetable}{cccccccc}
\vspace{0cm}
\tabletypesize{\scriptsize}
\tablecolumns{8}
\tablewidth{0pc}
\tablecaption{Relative Proper Motion\label{tab:rPM}}
\tablehead{
\colhead{Year} & \colhead{Source} & \colhead{Right Ascension} & \colhead{Declination} &  \colhead{$\Delta$(R.\,A.)} & \colhead{$\Delta$(Dec.)} & \colhead{Separation} & \colhead{Position Angle} \vspace{-0.5cm} \\
\colhead{} & \colhead{} & \colhead{(J2000)} & \colhead{(J2000)} & \colhead{(mas)} & \colhead{(mas)} & \colhead{(mas)} & \colhead{(mas)}  \vspace{-0.5cm} \\}
\startdata
1994.89 & A & 04$\rm ^h$31$\rm ^m$44$\rm ^s$.4975$\pm 0.0038$ & $+18^\circ$08$\farcm$31\farcs90$\pm0\farcs03$ & $$ & $$ & $$ & $$ \vspace{-0.3cm} \\
 $$        & B & 04$\rm ^h$31$\rm ^m$44$\rm ^s$.4625$\pm 0.0023$ & $+18^\circ$08$\farcm$32\farcs10$\pm0\farcs02$  & $-500.1\pm62.9$ & $200.0\pm35.3$ & $538.6 \pm 59.9$ & $291.8 \pm 4.3$ \\ 
2002.30 & A & 04$\rm ^h$31$\rm ^m$44$\rm ^s$.49701$\pm 0.00048$ & $+18^\circ$08$\farcm$31\farcs673$\pm0\farcs005$ & $$ & $$ & $$ & $$ \vspace{-0.3cm} \\
$$         & B & 04$\rm ^h$31$\rm ^m$44$\rm ^s$.46643$\pm 0.00088$ & $+18^\circ$08$\farcm$31\farcs864$\pm0\farcs009$ & $-435.9\pm14.4$ & $191.1\pm10.8$  & $475.9 \pm 13.9$ & $293.7 \pm 1.4$ \\ 
2012.91 & A & 04$\rm ^h$31$\rm ^m$44$\rm ^s$.506729$\pm 0.000053$ & $+18^\circ$08$\farcm$31\farcs4497$\pm0\farcs0010$ & $$ & $$ & $$ & $$ \vspace{-0.3cm} \\
$$         & B & 04$\rm ^h$31$\rm ^m$44$\rm ^s$.474554$\pm 0.000160$ & $+18^\circ$08$\farcm$31\farcs6744$\pm0\farcs0015$ & $-458.8\pm2.4$ & $224.7\pm1.9$ & $510.9 \pm 2.3$ & $296.1 \pm 0.2$  \\ 
\enddata
%\vspace{0.1cm}
%\tablecomments{}
\end{deluxetable}
\clearpage

%% This command is needed to show the entire author+affilation list when
%% the collaboration and author truncation commands are used.  It has to
%% go at the end of the manuscript.
%\allauthors

%% Include this line if you are using the \added, \replaced, \deleted
%% commands to see a summary list of all changes at the end of the article.
\listofchanges


\begin{thebibliography}{}
	\bibitem[ALMA Partnership et al.(2015)]{ALMA Partnership2015} ALMA Partnership et al. 2015, \apj, 808, L3
	\bibitem[Anglada et al.(2007)]{Anglada2007} Anglada, G., L\'opez, R., Estallela, R., Masegosa, J., Riera, A., \& Raga, A. C. 2007, \aj, 133, 2799
	\bibitem[Bate(2000)]{Bate2000} Bate, M. R. 2000, MNRAS, 314, 33
	\bibitem[Bate \& Bonnell(2005)]{Bate2005} Bate, M. R., \& Bonnell, I. A. 2005, MNRAS, 356, 1201
	\bibitem[Bate et al.(2002)]{Bate2002} Bate, M. R., Bonnell, I. A., \& Bromm, V. 2002, MNRAS, 332, L65
	\bibitem[Bate et al.(2003)]{Bate2003} Bate, M. R., Bonnell, I. A., \& Bromm, V. 2003, MNRAS, 339, 577
	\bibitem[Bertout et al.(1999)]{Bertout1999} Bertout, C., Robichon, N., \& Arenou, F. 1999, \aa, 352, 574
	\bibitem[Cha \& Whitworth(2003)]{Cha2003} Cha, S.-H., \& Whitworth, A. P. 2003, MNRAS, 340, 91
	\bibitem[Delgado-Donate et al.(2004a)]{DelgadoDonate2004a} Delgado-Donate, E. J., Clarke, C. J.,  Bate, M. R., \& Hodgkin, S. T.  2004a, MNRAS, 351, 617
	\bibitem[Delgado-Donate et al.(2004b)]{DelgadoDonate2004b} Delgado-Donate, E. J., Clarke, C. J., \& Bate, M. R. 2004b, MNRAS, 347, 759
	\bibitem[Galli \& Shu(1993a)]{Galli1993a} Galli, D., \& Shu, F. H. 1993a, ApJ, 417, 220
	\bibitem[Galli \& Shu(1993b)]{Galli1993b} Galli, D., \& Shu, F. H. 1993b, ApJ, 417, 243 
	\bibitem[Goldsmith et al.(2008)]{Goldsmith2008} Goldsmith, P. F., Heyer, M., Narayanan, G., Snell, R., Li, D., \& Brunt, C. 2006, \apj, 680, 428
	\bibitem[Goodwin et al.(2007)]{Goodwin2007} Goodwin, S. P., Kroupa, P., Goodman, A., \& Burkert, A. 2007, in Protostars andPlanets V, ed. B. Reipurth, D. Jewitt, \& K. Keil (Tucson, AZ: Univ. ArizonaPress), 133
	\bibitem[Goodwin et al.(2004a)]{Goodwin2004a} Goodwin, S. P., Whitworth, A. P., \& Ward-Thompson, D. 2004a, A\&A, 414, 633
	\bibitem[Goodwin et al.(2004b)]{Goodwin2004b} Goodwin, S. P., Whitworth, A. P., \& Ward-Thompson, D. 2004b, A\&A, 423, 169
	\bibitem[Goodwin et al.(2006)]{Goodwin2006} Goodwin, S. P., Whitworth, A. P., \& Ward-Thompson, D. 2006, A\&A, 452, 487
	\bibitem[Guilloteau et al.(2008)]{Guilloteau2008} Guilloteau, S., Dutrey, A., Pety, J., \& Gueth, F. 2008, A\&A, 478, L31
	\bibitem[Hayashi \& Pyo(2009)]{Hayashi2009} Hayashi, M. \& Pyo, T.-S. 2009, \apj, 694, 582
	\bibitem[Kenyon et al.(1994)]{Kenyon1994} Kenyon, S. J., Dobrzycka, D., \& Hartmann, L. 1994, \aj, 108, 1872
	\bibitem[Krist et al.(1999)]{Krist1999} Krist, J. E. et al. 1999, \apj, 515, L35
	\bibitem[Lim \& Takakuwa(2006)]{Lim2006} Lim, J., \& Takakuwa, S. 2006, \apj, 653, 425
	\bibitem[Lim et al.(2016)]{Lim2016} Lim, J., Yeung, P. K. H., Hanawa, T., Takakuwa, S., Matsumoto, T., \& Saigo, K. 2016, \apj, in press
	\bibitem[Lin et al.(2016)]{Lin2016} Lin, J. et al. 2016, arXiv:astro-ph/1605.03467v1
	\bibitem[L\'opez et al.(1995)]{Lopez1995} L\'opez, R. Raga, A., Riera, A., Anglada, G., \& Estallela, R. 1995, \mnras, 274, L19
	\bibitem[Lubow \& Ogilvie et al.(2000)]{Lubow2000} Lubow, S. H. \& Ogilvie, G. I. 2000, \apj, 538, 326
	\bibitem[Machida et al.(2008)]{Machida2008} Machida, M. N., Tomisaka, K., Matsumoto, T., \& Inutsuka, S. 2008, ApJ, 677, 327
%	\bibitem[Madlener et al.(2012)]{Madlener2012} Madlener, D., Wolf, S., Dutrey, A., \& Guilloteau, S. 2012, \aa, 543, 81
	\bibitem[Matsumoto \& Hanawa(2003)]{Matsumoto2003} Matsumoto, T., \& Hanawa, T. 2003, ApJ, 595, 913	
	\bibitem[Matsumoto et al.(2015)]{Matsumoto2015} Matsumoto, T., Onishi, T., Tokuda, K., \& Inutsuka, S. 2015, \mnras, 449, L123
	\bibitem[Mizuno et al.(1995)]{Mizuno1995} Mizuno, A., Onishi, T., Yonekura, Y., Nagahama, T., Ogawa, H. \& Fukui, Y. 1995, \apj, 445, L161
	\bibitem[Moriarty-Schieven et al.(2000)]{Moriarty-Schieven2000} Moriarty-Schieven, G. H., Powers, J. A., \& Butner, H. M., Wannier, P. G., \& Keene, J. 2000, \apj, 533, L143
	\bibitem[Moriarty-Schieven et al.(2006)]{Moriarty-Schieven2006} Moriarty-Schieven, G. H., Johnstone, D., Bally, J., \& Jenness, T. 2006, \apj, 645, 357
	\bibitem[Mundt et al.(1990)]{Mundt1990} Mundt, R., Ray, T. P., B\"uhrke, T., Raga, A. C., \& lf, J. 1990, \aap, 232, 37
	\bibitem[Peng et al.(2002)]{Peng2002} Peng, C. Y., Ho, L. C., Impey, C. D., \& Rix, H.-W. 2002, AJ, 124, 266
	\bibitem[Peng et al.(2010)]{Peng2010} Peng, C. Y., Ho, L. C., Impey, C. D., \& Rix, H.-W. 2010, AJ, 139, 2097
	\bibitem[P\'erez et al.(2012)]{Perez2012} P\'erez et al. 2012, \apj, 760, L17
	\bibitem[Pichardo et al.(2005)]{Pichardo2005} Pichardo, B., Sparke, L. S., \& Aguilar, L. A. 2005, MNRAS, 359, 521
	\bibitem[Reynolds(1986)]{Reynolds1986}Reyholds, S.~P. 1986, \apj, 304, 713
	\bibitem[Reipurth et al.(2000)]{Reipurth2000} Reipurth, B., Chan, K. C., Heathcote, S., Bally, J., \& Rodr\'iquez, L. F 2000, \aj, 120, 1449
	\bibitem[Reipurth et al.(2002)]{Reipurth2002} Reipurth, B., Rodr\'iquez, L. F., Anglada, G., \& Bally, J. 2002, \aj, 124, 1054
	\bibitem[Rodr\'iquez, Angalda, \& Raga(1995)]{Rodriquez1995} Rodr\'iquez, L. F., Anglada, G., \& Raga, A. 1995, \apj, 454, L149
	\bibitem[Segura-Cox(2016)]{Segura-Cox2016} Segura-Cox, D. M. et al. 2016, \apj, 817, L14
	\bibitem[Takakuwa et al.(2012)]{Takakuwa2012} Takakuwa, S., Saito, M., Lim, J., et al. 2012, \apj, 754, 52
	\bibitem[Takakuwa et al.(2013)]{Takakuwa2013} Takakuwa, S., Saito, M., Lim, J., \& Saigo, K. 2013, \apj, 776, 51
	\bibitem[Takakuwa et al.(2014)]{Takakuwa2014} Takakuwa, S., Saito, M., Saigo, K. et al. 2014, \apj, 776, 51
	\bibitem[Tobin et al.(2015)]{Tobin2015} Tobin, J. et al. 2015, \apj, 805, 125
\end{thebibliography}
\end{document}